\documentclass[galaxies,article,accept,oneauthor,pdftex,10pt,a4paper]{mdpi}
\pdfoutput=1

\usepackage{soul}
\usepackage{upgreek}
\usepackage{enumitem}
\usepackage{subfigure}
\usepackage{booktabs}
\usepackage{multirow}
\usepackage{longtable}
\usepackage{microtype}
\newcommand\myurl[1]{\changeurlcolor{black}\url{#1}\changeurlcolor{blue}}

\firstpage{1}
\articlenumber{x}
\doinum{10.3390/------}
\pubvolume{4}
\pubyear{2016}
\copyrightyear{2016}
\externaleditor{Academic Editor: Emilio Elizalde}
\history{Received: 20 October 2016; Accepted: 8 December 2016; Published: date} 

\usepackage{longtable} \usepackage{booktabs}
\usepackage{booktabs}
\usepackage{multirow}

\usepackage{tabularx}			
\newcolumntype{Y}{>{\centering\arraybackslash}X}
\newcolumntype{L}{>{\raggedright\arraybackslash}X}
\newcolumntype{R}{>{\raggedleft\arraybackslash}X}
\usepackage{upgreek}			
\usepackage{bm}					
\usepackage{physics}			
\usepackage{multirow}			
\usepackage{MnSymbol}			
\usepackage{datetime}			
\usepackage{longtable}			
\usepackage{xfrac}				
\usepackage{xspace}				
\usepackage[table]{xcolor}    	
\usepackage{adjustbox}			
\usepackage{array}				
\usepackage{dcolumn}			
\usepackage{empheq}				

\RequirePackage[l2tabu, orthodox]{nag}	
\newcolumntype{d}[1]{D{.}{.}{#1} }
\newcolumntype{O}[2]{%
    >{\adjustbox{angle=#1,lap=\width-(#2)}\bgroup}%
    l%
    <{\egroup}%
}
\newcolumntype{S}[1]{>{\centering\let\newline\\\arraybackslash\hspace{0pt}}m{#1}}

\newcommand{\f} {\fontfamily{ppl}\sfrac{7}{8}\xspace}	
\newcommand{\tf} {\fontfamily{ppl}\sfrac{3}{4}\xspace}	
\newcommand{\oha}{\fontfamily{ppl}\sfrac{1}{2}\xspace}	
\newcommand{\tha}{\fontfamily{ppl}\sfrac{3}{2}\xspace}	

\newcommand{\E}[1]{$\times 10^{#1}$}					
\newcommand{\vecp}{\vec{\phantom{p}}\hspace{-6pt}p}		

\newcommand{\kB}{k_{\textrm{B}}}				
\newcommand{\GF}{G_{\textrm{F}}}				
\newcommand{\GN}{G}							
\newcommand{\Gw}{G_\textrm{wk}}				

\newcommand{\g}{g_\star}			      		
\renewcommand{\ge}{g_{\star \epsilon}}			
\newcommand{\gn}{g_{\star n}}					
\newcommand{\gp}{g_{\star p}}					
\newcommand{\gs}{g_{\star s}}					

\renewcommand{\ggg}[1]{g_{\star #1}}			
\newcommand{\gee}[1]{g_{\star \epsilon #1}}		
\newcommand{\gnn}[1]{g_{\star n #1}}			
\newcommand{\gpp}[1]{g_{\star p #1}}			
\newcommand{\gss}[1]{g_{\star s #1}}			

\newcommand{\ee}{\epsilon}						
\newcommand{\nn}{n}								
\newcommand{\pp}{P}								
\renewcommand{\ss}{s}							

\Title{On Effective Degrees of Freedom in the Early~Universe}

\Author{Lars Husdal} 
\AuthorNames{Lars Husdal}

\address [1]{
Department of Physics, Norwegian University of Science and Technology, N-7491 Trondheim, Norway; lars.husdal@ntnu.no}

\abstract{We explore the effective degrees of freedom in the early Universe, from before the electroweak scale at a~few femtoseconds after the Big Bang until the last positrons disappeared a~few minutes later. We look at the established concepts of effective degrees of freedom for energy density, pressure, and entropy density, and introduce effective degrees of freedom for number density as well. We discuss what happens with particle species as their temperature cools down from  relativistic to semi- and non-relativistic temperatures, and then annihilates completely. This will affect the pressure and the entropy per particle. We also look at the transition from a~quark-gluon plasma~to a~hadron gas. Using a~list a~known hadrons, we use a~``cross-over'' temperature of 214 MeV, where the effective degrees of freedom for a~quark-gluon plasma~equals that of a~hadron gas.}


\begin{document}

\section{Introduction}
\label{Sec:Introduction}
The early Universe was filled with different particles. A~tiny fraction of a~second after the Big Bang, when the temperature was $10^{16}~\textrm{K} \approx 1~\textrm{TeV}$, all the particles in the Standard Model were present, and roughly in the same abundance. Moreover, the early Universe was in thermal equilibrium. At this time, essentially all the particles moved at velocities close to the speed of light. The average distance travelled and lifetime of these ultra-relativistic particles were very short. The frequent interactions led to the constant production and annihilation of particles, and as long as the creation rate equalled that of the annihilation rate for a~particle species, their abundance remained the same. The production of massive particles requires high energies, so when the Universe expanded and the temperature dropped, the~production rate of massive particles could not keep up with their annihilation rate. The heaviest particle we know about, the top quark and its antiparticle, started to disappear just one picosecond ($10^{-12}~\textrm{s}$) after the Big Bang. During the next minutes, essentially all the particle species except for photons and neutrinos vanished one by one. Only a~very tiny fraction of protons, neutrons, and electrons, what makes up all the matter in the Universe today, survived due to baryon asymmetry (the imbalance between matter and antimatter in the Universe). The fraction of matter compared to photons and neutrinos is less than one in a~billion, small enough to be disregarded in the grand scheme for the first stages of the Universe.

We know that the early Universe was close to thermal equilibrium from studying the Cosmic Microwave Background (CMB) radiation. Since its discovery in 1964 \cite{Penzias:1965}, the CMB has been thoroughly measured, most recently by the Planck satellite \cite{Planck:2015XIII}. After compensating for foreground effects, the CMB  almost perfectly fits that of a~black body spectrum, deviating by about one part in a hundred thousand~\cite{Trodden:2004}. It remained so until the neutrinos decoupled. For a~system in thermal equilibrium, we can use statistical mechanics to calculate quantities such as energy density, pressure, and entropy density. These quantities all depend on the number density of particles present at any given time. How~the different particles contribute to these quantities depends of their nature---most important being their mass and degeneracy. The complete contribution from all particles is a~result of the sum of all the particle species' effective degrees of freedom. We call these temperature-dependent functions $\g$, and~we have one for each quantity, such as $\gn$ related to number density, and $\ge$, $\gp$, and $\gs$, related~to energy density, pressure, and entropy density, respectively.

\begin{figure}[H]
\centering
	\includegraphics[width=0.75\textwidth]{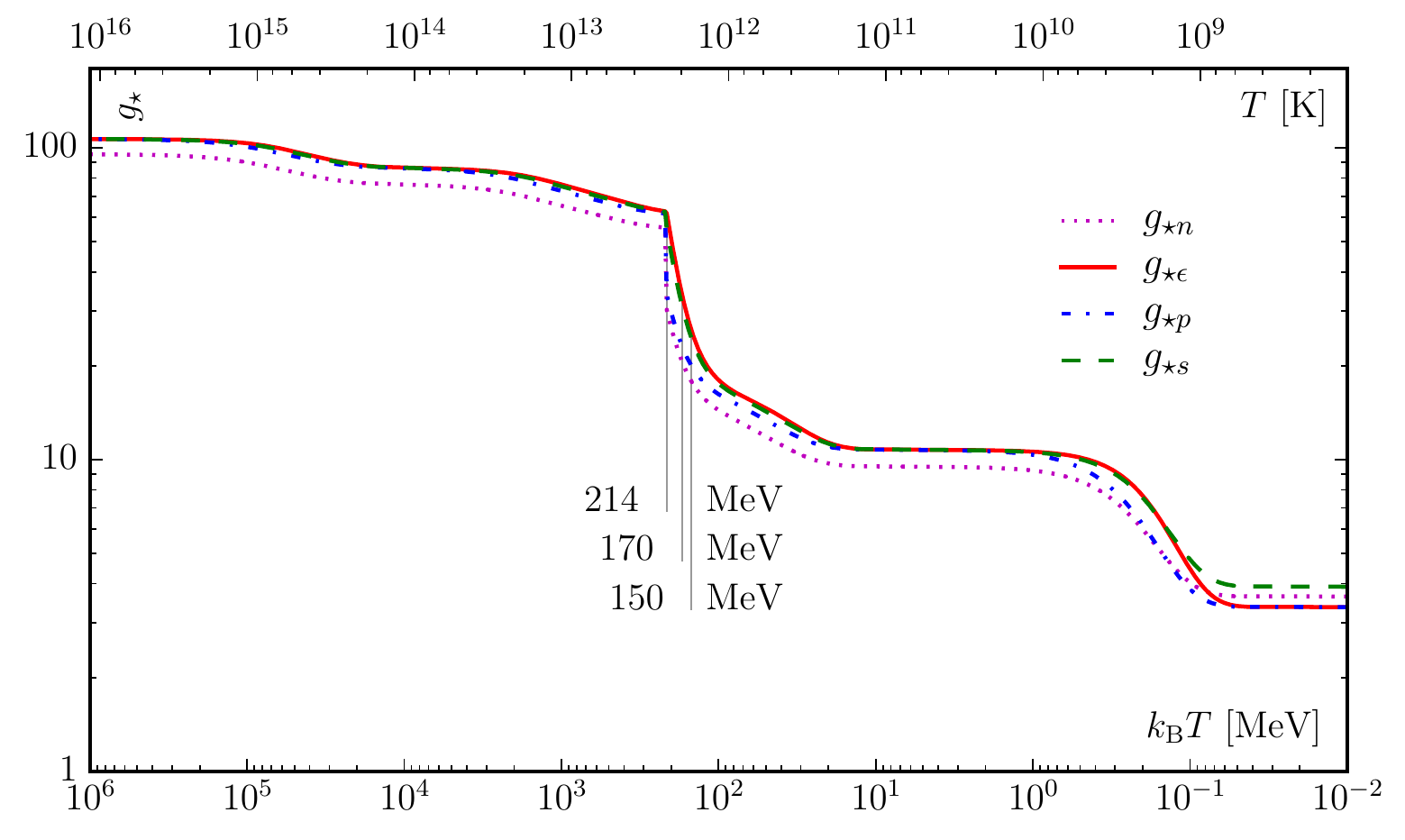}
	\caption{The evolution of the number density ($\gn$), energy density ($\ge$), pressure ($\gp$), and entropy density ($\gs$) as functions of temperature.}
	\label{Fig:gStarNRPS}
\end{figure}

In this paper, we will show how to calculate these four quantities ($\nn$, $\ee$, $\pp$, $\ss$), as well their associated effective degrees of freedom ($\gn$, $\ge$, $\gp$, $\gs$). These latter functions describe how the number of different particles evolve, and we have plotted these values in Figure~\ref{Fig:gStarNRPS}. Throughout this paper, we will look more closely at five topics.
After first having a~quick look at the elementary particles of the Standard Model and their degeneracy (Section \ref{Sec:ParticleDegeneracy}), we address the standard approach when everything is in thermal equilibrium in Section \ref{Sec:Theory}. Next, we take a~closer look at the behavior during the QCD phase transition; i.e., the transition from a~quark-gluon plasma~(QGP) to a~hot hadron gas (HG) in Section \ref{Sec:Evolution}. We then look at the behavior during neutrino decoupling (Section \ref{Sec:Decoupling}). For the fifth topic, we study how the temperature decreases as function of time (Section \ref{Sec:Time}). In Appendix \ref{Sec:GTable}, we have also included a~table with the values for all four $\g$s, as well as time, from temperatures of 10 TeV to 10 keV. The table includes three different transition temperatures as we go from a~QGP to a~HG. This article was inspired by the lecture notes by Baumann \cite{Baumann:Cosmology} and Kurki-Suonio \cite{KurkiSuonio:Cosmology}. Other important books on the subject are written by Weinberg \cite{Weinberg:Cosmology, Weinberg:GravitationAndCosmology}, Kolb and Turner \cite{Kolb:EarlyUniverse}, Dodelson \cite{Dodelson:ModernCosmology}, Ryden \cite{Ryden:IntroductionToCosmology}, and Lesgourgues, Mangano, Miele, and Pastor \cite{Lesgourgues:NeutrinoCosmology}.

\section{Notations and Conventions}
\label{Sec:Notations}
The effective degrees of freedom of a~particle species is defined  relative to the photon. This~is not just an arbitrary choice, but chosen since the photon is massless, and whose density history is best known. The most important source of information about the early Universe comes from the CMB photons. Even though the photon is the natural choice as a~reference particle, technically any particle could be used. Additionally, when talking about effective degrees of freedom, we most often do so in the context of energy density $\ge$, which in most textbooks is just called ``$\g$''. Here, we use the notation $\ge$ for that matter, and $\g$ as a~collective term for all four quantities.

The term ``particle annihilations'' frequently appears in this paper. Strictly speaking, we have particle creations and annihilations all the time, but in this context, ``particle annihilations'' refers to periods where the annihilation rate is (noticeable) faster than the production rate for a~particle species.

In many textbooks, the value of the speed of light ($c$), the Boltzmann constant ($\kB$), and the Planck reduced constant ($\hbar$) are set to unity. We have chosen to keep these units in our equations to avoid any problems with dimensional analysis during actual calculations. One of the advantages of using $\hbar=c=\kB=1$ is that we can use temperature, energy, and mass interchangeably. For our equations, we use $\kB T$ and $mc^2$ when we want to express temperature and mass in units of MeV, but in the main text, when we talk about temperature and mass, it is implied that these are $\kB T$ and $mc^2$.

Simplifications are important when we first want to approach a~new subject. One of our assumptions in this paper is that the early Universe was in total thermal equilibrium. There were, however, periods where this was not so. In those cases, viscous effects drove the system (the Universe) towards equilibrium. This increased the entropy. For our purposes, all viscous effects have been neglected. Some relevant papers address this issue \cite{Husdal:2016Viscosity, Hoogeveen:1986, Caderni:1977}.

\section{The Standard Model Particles and Their Degeneracy}
\label{Sec:ParticleDegeneracy}

\begin{figure}[H]
	\centering
	\includegraphics[width=0.70\textwidth]{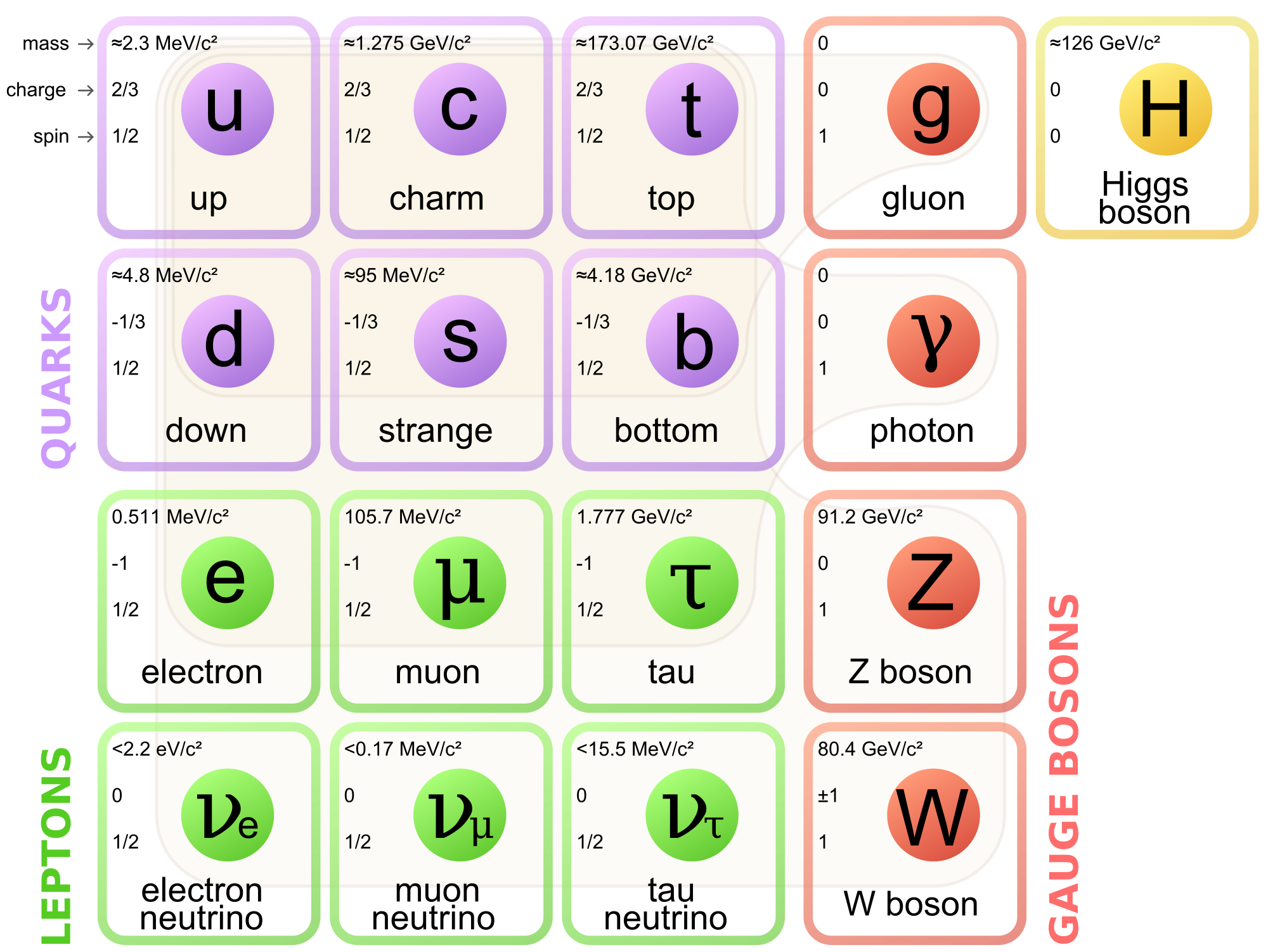}
	\caption{All particle species of the Standard Model of elementary particles.}
	\label{Fig:StandardModel}
\end{figure}

Let us start by looking at the degeneracy of the different particle species---their intrinsic degrees of freedom, $g$. The Standard Model of elementary particles are often displayed as in Figure \ref{Fig:StandardModel}. The~quarks, leptons, and neutrinos are grouped into three families, shown as the first three columns. These are all fermions. The two last columns are the bosons. The fourth column consists of force mediator particles, also called gauge bosons. These are the eight gluons, the photon, and the three massive gauge bosons. The Higgs boson that was discovered at CERN in 2012 \cite{HiggsDiscoveryATLAS, HiggsDiscoveryCMS} comprises the fifth and last column.

A~particle's degeneracy depends on its nature and which properties it possesses. We have listed these as four different columns in Table \ref{Tab:DegreesOfFreedom}. They are:
(1) Number of different flavors. These are different types of particles with similar properties, but different masses. These are listed as separate entries in Figure \ref{Fig:StandardModel};
(2) Existence of antiparticles. Antiparticles have different charge, chirality, and color than their particle companion. Not all particles have anti-partners (e.g., the photon);
(3) Number of color states. Strongly interacting particles have color charge. For quarks and their anti-partners, there are three possibilities (red, green, blue, or antired, antigreen, antiblue). Gluons have eight possible color states. These are superpositions of combined states of the three plus three colors;
(4)~Number~of possible spin states. We remember from quantum mechanics that all bosons have integer spins, while~fermions have half integer spins, both in units of $\hbar$. The spin alignment of a particle in some direction is~called its polarization. Quarks and the charged leptons have two possible polarizations: $+$\oha~or $-$\oha. Another way of saying this is that they can be either
~left-handed or right-handed. Neutrinos, on the other hand, can only be left-handed (and antineutrinos only right-handed), so they only have one spin state. Actually, whether neutrinos are Dirac or Majorana~fermions is still an open question. Majorana~fermions are their own antiparticles, while Dirac fermions have distinct particles and antiparticles. In the latter case, we expect there to be additional right-handed neutrinos and left-handed antineutrinos, whose weak interaction is suppressed. These ``new'' neutrinos are expected to have negligible density compared to the left-handed neutrinos and right-handed antineutrinos \cite{Lesgourgues:2012}. The book by Lesgourgues, Mangano, Miele, and Pastor \cite{Lesgourgues:NeutrinoCosmology} also discusses this topic in detail. The massive spin-1 bosons ($\textrm{W}^{\pm}$ and $\textrm{Z}^{0}$) have three possible polarizations ($-1,0,1$): one longitudinal and two transverse. The massless spin-1 bosons (photons and gluons) have only two possible polarizations, namely the transverse ones.
The Higgs particle is a~scalar particle and has spin-0. Finally, we should say that hadrons can have multiple possible spin states, depending on their composition.

\begin{table}[H]
\small
\centering
	\caption{The Standard Model of elementary particles and their degeneracies.}
	\label{Tab:DegreesOfFreedom}
\begin{tabular}{lccccc}
\toprule
	& \textbf{Flavors}	& \textbf{Particle + Antiparticle}		& \textbf{Colors} 	& \textbf{Spins}	& \textbf{Total} \\\midrule
Quarks (u,\,d,\,c,\,s,\,t,\,b)					 	  & 6 &  2  & 3 & 2 & 72 \\
Charged leptons (e, $\upmu$, $\uptau$)			  	  & 3 &  2  & 1 & 2 & 12 \\
Neutrinos ($\nu_\textrm{e}$, $\nu_\upmu$, $\nu_\uptau$) & 3 &  2  & 1 & 1 &  6 \\
Gluons (g)										  	  & 1 &  1  & 8 & 2 & 16 \\	
Photon  ($\gamma$)									  	  & 1 &  1  & 1 & 2 &  2 \\
Massive gauge bosons (W$^\pm$, Z$^0$)				  	  & 2 & 2, 1& 1 & 3 &  9 \\	
Higgs bosons (H$^0$)								  	  & 1 &  1  & 1 & 1 &  1 \\	
All elementary particles								  	  &17 &     &   &   &118 \\
\bottomrule
\end{tabular}
\end{table}

At high temperatures where all the particles of the Standard Model are present, we have 28~bosonic and 90 fermionic degrees of freedom. It turns out that fermions do not contribute as much as bosons, since they can not occupy the same state. We will get back to this in the next section, and just say that fermions have $28 + \textrm{\f} \times 90 = 106.75$ effective degrees of freedom for energy density, pressure, and entropy density. For the number density, the effective degrees of freedom is $28 + \textrm{\tf} \times 90 = 95.5$.

\section{Statistical Mechanics of Ideal Quantum Gases in Thermodynamic Equilibrium}
\label{Sec:Theory}
In this section, we briefly review the statistical mechanics of ideal quantum gases in thermal equilibrium. We also introduce the concept of effective number of degrees of freedom for a~particle species, and how to count these as functions of the temperature.

\subsection{Thermodynamic Functions}
\label{SubSec_StatMech}
In order to calculate the thermodynamic functions, we need to know the single-particle energies of the system. We consider a~cubic box with periodic boundary conditions, and with sides of length $L$ and volume $V = L^3$. Solving the Schr\"odinger equation for a~particle, we find the possible momentum~eigenvalues
\begin{equation}
  \vecp = \frac{h}{L}(n_1 \vec{e}_x + n_2 \vec{e}_y + n_3 \vec{e}_z) \;,
  \label{Eq:ScrodingerPBC}
  \end{equation}
where $h$ is the Planck constant, $n_i=$ 0, $\pm$1, $\pm$2, $\pm$3, ..., and $\vec{e}_x$, $\vec{e}_y$, $\vec{e}_z$ are the standard units vectors in three-dimensional Euclidean space.  The energy of a~particle with mass $m$ and momentum $\vecp$ is $E(\vecp)=\sqrt{m^2c^4+\vecp^2c^2}$.

In thermal equilibrium, the probability that a~single-particle state with momentum $\vecp$ and energy $E(\vecp)$ is occupied is given by the Bose--Einstein or Fermi--Dirac distribution functions
\begin{equation}
  f(\vecp) = \frac{1}{e^{(E(\vec{p})-\mu) / (\kB T)} \pm 1} \;,
  \label{Eq:DistributionFunction}
  \end{equation}
where the upper sign is for fermions and the lower sign for bosons. Moreover, $\kB$ is the Boltzmann constant and $\mu$ is the chemical potential. In order to find the total number of particles occupying a~state with energy $E$, we must find the density of states in phase space. We see from Equation (\ref{Eq:ScrodingerPBC}) that the number of possible states in momentum space is $L^3/h^3$. By dividing by the volume, $L^3$, as well, we are left with the factor $(1/h)^3$. If there is an additional degeneracy $g$ (for example, spin), we can write the density of states (dos) as
\begin{equation}
  \textrm{dos} = \frac{g}{h^3} = \frac{g}{(2\pi)^3 \hbar^3} \;.
  \end{equation}

The density of particles with momentum $\vecp$ is then given by
\begin{equation}
  n(\vecp) = \frac{g}{(2\pi)^3 \hbar^3} \times f(\vecp)\;.
  \end{equation}

The total density of particles, $n$, can then be written as an integral over three-momentum involving the distribution function as
\begin{equation}
  n = \frac{g}{(2\pi)^3 \hbar^3} \int f(\vecp) \dd[3]{\vecp} \;.
  \end{equation}

By multiplying the distribution function (\ref{Eq:DistributionFunction}) with the energy and integrating over three-momentum, we obtain the energy density $\ee$ of the system. The pressure, $\pp$, can be found in a~similar manner by multiplying the distribution function with $|\vecp|^2/(3E/c^2)$ (a~nice derivation of this is shown by Baumann \cite{Baumann:Cosmology}). This yields the integrals
\begin{align}
  \ee &= \frac{g}{(2\pi)^3 \hbar^3}
			   \int E(\vecp) f(\vecp) \dd[3]{\vecp} \;,\\
  \pp &= \frac{g}{(2\pi)^3 \hbar^3}
			   \int \frac{|\vecp|^2}{3(E/c^2)} f(\vecp) \dd[3]{\vecp} \;.
  \end{align}

Finally, let us mention the entropy density $\ss$. It can be calculated from the thermodynamic relation
\begin{equation}
  \ss = \frac{\ee + \pp - \mu_\textrm{T}}{T} \;,
  \label{Eq:Entropy}
  \end{equation}
where the index $\mu_\textrm{T}$ is the total chemical potential. We will get back to chemical potentials in Section \ref{Sec:ChemicalPotential}.

\subsection{From Momentum to Energy Integrals}
It is sometimes more convenient to use energy, $E$, instead of the momentum, $\vecp$, as the integration variable. By integrating over all angles, we can replace $\dd[3]{\vecp} $ by $4\pi|\vecp|^2 \dd{\vecp}$. Using the energy momentum relation, we find $|\vecp|=\sqrt{E^2-m^2c^2}/c$ and $c\vecp\dd{\vecp}=E\dd{E}$. We can simplify these formulas further by introducing the dimensionless variables $u$, $z$, and $\hat{\mu}$.
\begin{equation}
  u=\frac{E}{\kB T} \;, 	\qquad
  z=\frac{mc^2}{\kB T} \;, 	\qquad
  \hat{\mu}=\frac{\mu}{\kB T} \;.
  \label{defrat}
  \end{equation}

This yields the following expressions for the number density, energy density, and pressure for a~species $j$, and for all species (as this is simply the sum of all particle species).
\vspace{12pt}
\begin{subequations}
\begin{align}
  \nn_j(T)	&= \frac{g_j}{2 \pi^2 \hbar^3} \int_{m_jc^2}^\infty
			   \frac{E \sqrt{E^2-m_j^2c^4}}{e^{(E-\mu_j)/\kB T} \pm 1} \dd{E} \\
  			&= \frac{g_j}{2\pi^2} \left( \frac{\kB T}{\hbar c} \right)^3
			   \int_{z_j}^\infty \frac{u\sqrt{u^2-z_j^2}}{e^{u-\hat{\mu}_j} \pm 1} \dd{u} \;, \\
  \nn(T)	 = \sum_j \nn_j
  			&= \sum_j \frac{g_j}{2\pi^2} \left( \frac{\kB T}{\hbar c} \right)^3
			   \int_{z_j}^\infty \frac{u\sqrt{u^2-z_j^2}}{e^{u-\hat{\mu}_j} \pm 1} \dd{u} \;,
  \end{align}
  \label{Eq:nj}
\end{subequations}
\vspace{-18pt}
\begin{subequations}
\begin{align}
  \ee_j(T) 	&= \frac{g_j}{2 \pi^2 \hbar^3} \int_{m_jc^2}^\infty
		       \frac{E^2 \sqrt{E^2-m_j^2c^4}}{e^{(E-\mu_j)/\kB T} \pm 1} \dd{E} \\
  			&= \frac{g_j}{2\pi^2} \frac{(\kB T)^4}{(\hbar c)^3} \int_{z_j}^\infty
			   \frac{u^2\sqrt{u^2-z_j^2}}{e^{u-\hat{\mu}_j} \pm 1} \dd{u} \;, \\
  \ee(T)	 = \sum_j \ee_j
  			&= \sum_j \frac{g_j}{2\pi^2} \frac{(\kB T)^4}{(\hbar c)^3} \int_{z_j}^\infty
			   \frac{u^2\sqrt{u^2-z_j^2}}{e^{u-\hat{\mu}_j} \pm 1} \dd{u} \;,
  \end{align}
  \label{Eq:ej}
  \end{subequations}
  \vspace{-18pt}
\begin{subequations}
\begin{align}
  \pp_j(T) 	&= \frac{g_j}{6 \pi^2 \hbar^3} \int_{m_jc^2}^\infty
			   \frac{(E^2-m_j^2c^4)^{3/2}}{e^{(E-\mu_j)/\kB T} \pm 1} \dd{E} \\
  			&= \frac{g_j}{6\pi^2} \frac{(\kB T)^4}{(\hbar c)^3} \int_{z_j}^\infty
  			   \frac{(u^2-z_j^2)^{3/2}}{e^{u-\hat{\mu}_j} \pm 1} \dd{u} \;, \\
  \pp(T)	 = \sum_j \pp_j
  			&= \sum_j \frac{g_j}{6\pi^2} \frac{(\kB T)^4}{(\hbar c)^3} \int_{z_j}^\infty
  			   \frac{(u^2-z_j^2)^{3/2}}{e^{u-\hat{\mu}_j} \pm 1} \dd{u} \;.
  \end{align}
  \label{Eq:pj}
\end{subequations}

As shown in Equation (\ref{Eq:Entropy}) we can find the entropy density for a~single species $j$ and the total entropy as:
\begin{subequations}
\begin{align}
  \ss_j(T) 	&= \frac{\ee_j + \pp_j - \mu_j n_j}{T} \; , \\
  \ss(T)     = \sum_j \ss_j
  			&= \sum_j \frac{\ee_j + \pp_j - \mu_j n_j}{T}
  			 = \frac{\ee + \pp - \sum_j\mu_j n_j}{T} \; .
  \end{align}
  \label{Eq:sj}
\end{subequations}

\subsection{Chemical Potentials}
\label{Sec:ChemicalPotential}
Before we proceed, we briefly discuss the chemical potentials. We recall from statistical mechanics that we can introduce a~chemical potential $\mu_j$ for each conserved charge $Q_j$. This is done by replacing the Hamiltonian $H$ of the system with $H-\mu_jN_{Q_j}$, where $N_{Q_j}$ is the number operator of particles with charge $Q_j$.

In the Standard Model, there are five independent conserved charges. These are electric charge, baryon number, electron-lepton number, muon-lepton number, and tau-lepton number. This means there are also five independent chemical potentials \cite{Weinberg:GravitationAndCosmology}. The chemical potentials are determined by the number densities. The electric charge density is very close to zero. The baryon density is estimated to be less than a~billionth of the photon density \cite{Bennett:2013, PDG:2014}. Lepton density is also thought to be very small, on the same order as the baryon number. According to Weinberg \cite{Weinberg:GravitationAndCosmology}, for an~early Universe scenario, we can put all these numbers equal to zero to a~good approximation. For a~correct representation of the Universe, the chemical potentials cannot all cancel out---otherwise, there would be no matter present today. For more general calculations including chemical potentials, the book by  Weinberg is recommended \cite{Weinberg:Cosmology}. The implications of a~large neutrino chemical potential is discussed by Pastor and Lesgourgues \cite{Lesgourgues:1999}. Mangano, Miele, Pastor, Pisanti, and Sarikasa~discuss the chemical potentials and their influence on the effective number of neutrino species \cite{Mangano:2011} (we will briefly mention effective neutrino species in Section \ref{SubSec:NeutrinoTemperature}).

\subsection{Massless Particle Contributions}
\label{Sec:MasslessParticleContributions}
In Equations (\ref{Eq:nj})--(\ref{Eq:pj}), we see how dimensionless units, $u$, $z$, and $\hat{\mu}$, simplifies the integrals. In~the ultrarelativistic limit, we can ignore the particle masses. Moreover, as we have set the chemical potentials to zero, we can easily solve the dimensionless integrals appearing in Equations~(\ref{Eq:nj}b), (\ref{Eq:ej}b), and (\ref{Eq:pj}b) analytically. Since the integrals for energy density and pressure in the massless cases are the same, we find:
\begin{equation}
  \int_0^\infty \frac{u^2}{e^u \pm 1} \dd{u}
  = \begin{cases}
   		\frac{3}{2} \zeta~(3) \simeq 1.803\\
   		2   \zeta~(3)		\simeq 2.404
    \end{cases}
  \begin{array}{l}
	\phantom{\frac{\pi^4}{15}} \textrm{(Fermions)} \;, \\
    \phantom{\frac{\pi^4}{15}} \textrm{(Bosons)} \;,
    \end{array}
  \label{Eq:nIntegrals}
  \end{equation}
\begin{equation}
  \int_0^\infty \frac{u^3}{e^u \pm 1} \dd{u}
  = \begin{cases}
		\frac{7}{8} \frac{\pi^4}{15} \simeq 5.682 \\
   		\frac{\pi^4}{15}  			 \simeq 6.494
    \end{cases}
  \begin{array}{l}
	\phantom{\frac{\pi^4}{15}} \textrm{(Fermions)} \;, \\
    \phantom{\frac{\pi^4}{15}} \textrm{(Bosons)} \;,
    \end{array}
  \label{Eq:epIntegrals}
  \end{equation}
where $\zeta(3)$ is the Riemann zeta~function of argument $3$. Using these results, we find the values for $\nn$, $\ee$, $\pp$, and indirectly $s$ for massless bosons and fermions:
\begin{align}
  &\nn_\textrm{b}(T)    = g \frac{\zeta(3)}{\pi^2} \frac{(\kB T)^3}{(\hbar c)^3} \;,
  &&\nn_\textrm{f}\,(T) = \frac{3}{4} g \frac{\zeta(3)}{\pi^2} \frac{(\kB T)^3}{(\hbar c)^3} \;, \\
  &\ee_\textrm{b}(T)    = g \frac{\pi^2}{30} \frac{(\kB T)^4}{(\hbar c)^3} \;,
  &&\ee_\textrm{f}\,(T) = \frac{7}{8} g \frac{\pi^2}{30} \frac{(\kB T)^4}{(\hbar c)^3} \;, \\
  &\pp_\textrm{b}(T)    = g \frac{\pi^2}{90} \frac{(\kB T)^4}{(\hbar c)^3} \;,
  &&\pp_\textrm{f}\,(T) = \frac{7}{8} g \frac{\pi^2}{90} \frac{(\kB T)^4}{(\hbar c)^3} \;, \\
  &\ss_\textrm{b}(T)    = g \frac{2\pi^2}{45} \frac{\kB^{\;4} T^3}{(\hbar c)^3} \;,
  &&\ss_\textrm{f}\,(T) = \frac{7}{8} g \frac{2\pi^2}{45} \frac{\kB^{\;4} T^3}{(\hbar c)^3} \;.
  \label{Eq:nepBF}
  \end{align}

Here the subscript b is for bosons, and f is for fermions. We see that solving the integrals gives a~difference between fermions and bosons, namely a~factor \tf for the number density and \f for energy density and pressure. We will call these two factors the ``fermion prefactors''. We also see that the pressure is simply one third that of the energy density, while the entropy density can be found by multiplying the energy density~by~$4/(3T)$.

\subsection{Effective Degrees of Freedom}
\label{Sec:EffectiveDegreesOfFreedom}

In most cases we cannot ignore the particle masses. In these cases, we must solve the integrals in Equations~(\ref{Eq:nj}b), (\ref{Eq:ej}b), and (\ref{Eq:pj}b) numerically. The integrals are decreasing functions of the temperature, and they vanish in the limit $\kB T/mc^2 \rightarrow 0$. We can normalize these by dividing their values by the case of the photon (but with $g$ equal to one). As we recall,the photon has a~bosonic nature with $m=0$ and $\mu=0$.
This means that for massive particles at high temperature ($\kB T \gg mc^2$), one actual degree of freedom for bosons contributes as much as~one degree of freedom for photons, and the fermions a~little less. As the temperature drops, and less particles are created, the effective contributions will be smaller. By including the intrinsic degrees of freedom ($g$), we find each particle species' \textit{effective degree of freedom}, $\ggg{_j}$:
\begin{align}
  \gnn{_j}(T) =
  \frac{\frac{g_j}{2\pi^2} \left( \frac{\kB T}{\hbar c} \right)^3 }
       {\frac{1  }{2\pi^2} \left( \frac{\kB T}{\hbar c} \right)^3 }
  \frac{\int_{z_j}^\infty \frac{u\sqrt{u^2-z_j^2}}{e^u \pm 1} \dd{u} }
       {\int_{0  }^\infty \frac{u^2              }{e^u \pm 1} \dd{u}}
  &= \frac{g_j}{2 \zeta(3)}
  \int_{z_j}^\infty \frac{u\sqrt{u^2-z_j^2}}{e^u \pm 1} \dd{u} \;,
  \label{Eq:gStarnj}
   \end{align}
\vspace{12pt}
   \begin{align}
  \gee{_j}(T) =
  \frac{\frac{g_j}{2\pi^2} \frac{(\kB T)^4}{(\hbar c)^3} }
       {\frac{1  }{2\pi^2} \frac{(\kB T)^4}{(\hbar c)^3} }
  \frac{\int_{z_j}^\infty \frac{u^2\sqrt{u^2-z_j^2}}{e^u \pm 1} \dd{u} }
       {\int_{0  }^\infty \frac{u^3                }{e^u \pm 1} \dd{u} }
   &= \frac{15 g_j}{\pi^4}
  \int_{z_j}^\infty \frac{u^2\sqrt{u^2-z_j^2}}{e^u \pm 1} \dd{u} \;,
  \label{Eq:gStarej}\\
  \gpp{_j}(T) =
  \frac{\frac{g_j}{6\pi^2} \frac{(\kB T)^4}{(\hbar c)^3} }
       {\frac{1  }{6\pi^2} \frac{(\kB T)^4}{(\hbar c)^3} }
  \frac{\int_{z_j}^\infty \frac{(u^2-z_j^2)^{3/2}}{e^u \pm 1} \dd{u} }
       {\int_{0  }^\infty \frac{ u^3             }{e^u \pm 1} \dd{u} }
  &= \frac{15 g_j}{\pi^4}
    \int_{z_j}^\infty \frac{(u^2-z_j^2)^{3/2}}{e^u \pm 1} \dd{u} \;,
  \label{Eq:gStarpj}\\
  \gss{_j}(T) &= \frac{3\gee{_j}(T) + \gpp{_j}(T)}{4}  \;.
  \label{Eq:gStarsj}
  \end{align}

\begin{figure}[H]
	\centering
\includegraphics[width=0.98\textwidth]{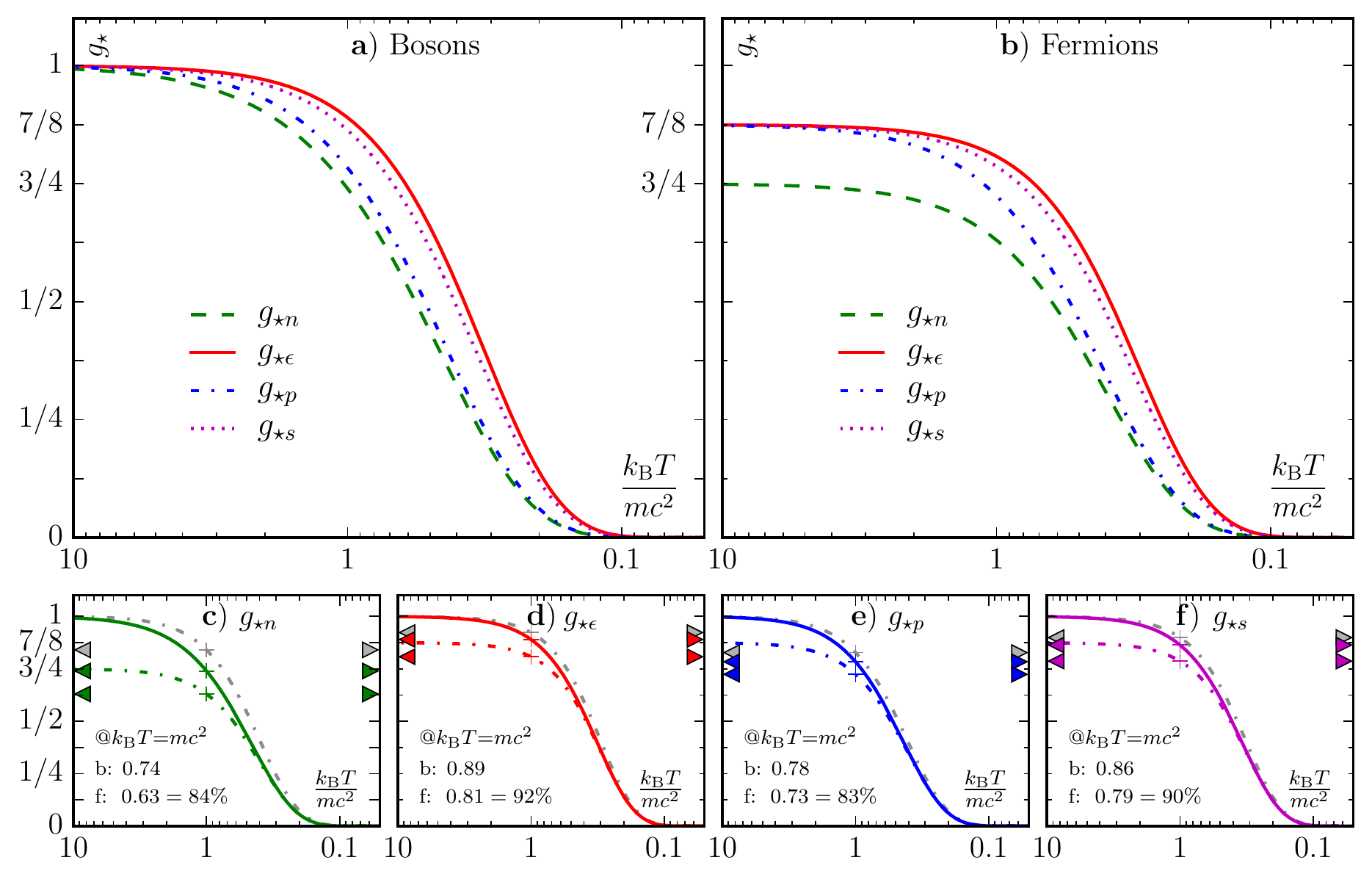}
	\caption{The 
	effective degrees of freedom $\gn$, $\ge$, $\gp$, and $\gs$ for bosons (\textbf{a}) and fermions (\textbf{b}) per intrinsic degree of freedom. A~more detailed look at each of the four $\g$s is given in the lower four panels (\textbf{c}--\textbf{f}). Here the solid colored curves are for the bosons, and the dash-dotted colored curves are for the fermions. The grey dash-dotted curves represent the fermions' contribution compared to its own relativistic value (such that it is 100\% for $T \rightarrow \infty$). We~have included the relative values at $\kB T =  mc^2$ for the four cases (marked with ``+'' symbols). During~particle annihilations, the energy density falls slower than the other quantities due to the impact of the rest mass energy. At temperatures close to the rest mass of some massive particle species, this rest mass is substantial to their total energy.}
	\label{Fig:BosonFermionIntegrals}
\end{figure}

In Figure~\ref{Fig:BosonFermionIntegrals}, we have plotted the effective degrees of freedom for massive bosons (panel \textbf{a}) and fermions (panel \textbf{b}) with $g=1$ (and $\mu=0$) as functions of the temperature. We~have also listed the results in Table \ref{Tab:OneParticleContribution} in Appendix
\ref{Sec:OneParticleContribution}. When the temperature is equal to the mass ($\kB T=mc^2$), the~effective degrees of freedom for energy density is approximately 0.9 for bosons and 0.8 for fermions, compared to that of the photon. For number density, pressure, and entropy density, they are a~little~lower.

The effective degrees of freedom are defined as functions of the corresponding variables and~temperature. We find the total effective degrees of freedom for $\gn$, $\ge$, and $\gp$ by summing Equations~(\ref{Eq:gStarnj})--(\ref{Eq:gStarsj}) over all particle species $j$:
\begin{subequations}
\begin{align}
  \gn (T) &\equiv \frac{\pi^2}{\zeta(3)} \frac {\nn(T)}{T^3}
  \label{Eq:gStarNumberDensityDefined} \\
  &= \sum_j \frac{g_j}{2 \zeta(3)}
  \int_{z_j}^\infty \frac{u\sqrt{u^2-z_j^2}}{e^u \pm 1} \;,
  \label{Eq:gStarNumberDensityThEq}
  \end{align}
\end{subequations}
\vspace{-16pt}
\begin{subequations}
\begin{align}
  \ge (T) &\equiv \frac{30}{\pi^2} \frac{\ee(T)}{T^4}
  \label{Eq:gStarEnergyDensityDefined}\\
  &= \sum_j \frac{15g_j}{\pi^4} \int_{z_j}^\infty \frac{u^2\sqrt{u^2-z_j^2}}{e^u \pm 1} \dd{u} \;,
  \label{Eq:gStarEnergyDensityThEq}
  \end{align}
\end{subequations}
\vspace{-16pt}
\begin{subequations}
\begin{align}
  \gp (T) &\equiv \frac{90}{\pi^2} \frac{\pp(T)}{T^4}
  \label{Eq:gStarPressureDefined}\\
  		&=  \sum_j \frac{15g_j}{\pi^4}
  			\int_{z_j}^\infty \frac{(u^2-z_j^2)^{3/2}}{e^u \pm 1} \dd{u} \;.
  \label{Eq:gStarPressureThEq}
  \end{align}
\label{Eq:gStarAll3Defined}
\end{subequations}

Finally, the effective degrees of freedom associated with entropy is then:
\begin{subequations}
\begin{align}
  \gs  (T) &\equiv \frac{45}{2\pi^2} \frac{\ss(T)}{T^3}
  \label{Eq:gStarEntropyDensityDefined} \\
  	  &= \frac{3\ge(T) + \gp(T)}{4} \;.
  \label{Eq:gStarEntropyDensityThEq}
  \end{align}
\end{subequations}

We again emphasize that Equations (\ref{Eq:gStarNumberDensityThEq}), (\ref{Eq:gStarEnergyDensityThEq}), (\ref{Eq:gStarPressureThEq}), and (\ref{Eq:gStarEntropyDensityThEq}) are only valid for a~system in thermal equilibrium (i.e., all the particles have the same temperature). It turns out that after the neutrinos decouple from the electromagnetically interacting particles (i.e., photons, electrons, and positrons) and the electrons and positrons annihilate, we cannot calculate the four $\g$s that straightforwardly. We will return to neutrino decoupling in Section \ref{SubSec:NeutrinoDecoupling}.

\section{Particle Evolution During the Cooling of the Universe}
\label{Sec:Evolution}

Our analysis starts with all the particles of the Standard Model present. As the Universe expands and cools, the annihilation rate of the more massive particles will become smaller and smaller compared to their creation rate. As the heavier particles disappear, this again will lead to a~relatively larger creation rate for all the remaining lighter particle species. The overall number of particles in a~comoving volume will thus remain (almost) constant. A~few minutes after the Big Bang, when the temperature was down to 10 keV (corresponding to 100 million Kelvin), the Universe was mainly filled with photons and  neutrinos. As we mentioned in Sections \ref{Sec:Introduction} and \ref{Sec:ChemicalPotential}, a~small---and at this stage, negligible---portion of matter survived due to the baryon asymmetry. Without the presence of antiparticles, the matter particles (i.e.,~nucleons~and electrons) thus survived and ``froze out'' when their reaction rate (i.e.,~annihilation~and creation rate) became slower than the expantion rate of the Universe (or equivalently, when the time scale of the weak interaction became longer than the age of the Universe) ~\cite{Kolb:EarlyUniverse}. This process has some similarities with the decoupling of the neutrinos (which we will discuss in more detail in Section \ref{Sec:Decoupling}). These relic matter particles still interact with the photons and remain in thermal equilibrium until after the photon decoupling at around 380,000 years after the Big Bang~\cite{Planck:2015XIII}. Although negligible in the early stages of the Universe, matter eventually became the dominant energy contributor around 47,000 years after the Big Bang \cite{Ryden:IntroductionToCosmology}. This is because non-relativistic (cold) matter receive their energy mainly from their rest mass. The energy density for cold matter goes as $T^{-3}$. This is solely due to the dilution of the particles. The~kinetic contribution to the energy is negligible. Radiation (massless particles) goes as $T^{-4}$, because it is also subject to redshift as the Universe expands. A~simple overview of events which affects the four $\g$s is given in Table \ref{Tab:gstar}.
\begin{table}[H]
\small
	\centering
	\caption{List of events which impacts $\gn$, $\ge$, $\gp$, and $\gs$. For the particle annihilation events, we have here used the particle masses as a~reference. By combining this Table with Table \ref{Tab:OneParticleContribution} in Appendix~\ref{Sec:OneParticleContribution}, we get a~more precise picture.}
	\label{Tab:gstar}
	\begin{tabular}{lccccc}
		\toprule
\textbf{Event} & \textbf{Temperature} & $\bm{\gn}$ & $\bm{\ge}$ & $\bm{\gp}$ & $\bm{\gs}$\\ \midrule
		&& 95.5 & 106.75 & 106.75 & 106.75  \\
Annihilation of ${\mathrm{t \bar{t}}}$ quarks & <173.3 GeV$$$$ \\
		&& 86.5 & 96.25 & 96.25 & 96.25 \\
Annihilation of Higgs boson & <125.6 GeV$$$$ \\
		&& 85.5 & 95.25 & 95.25 & 95.25 \\
Annihilation of Z$^0$ boson & <91.2 GeV $$$$ \\
		&& 82.5 & 92.25 & 92.25 & 92.25 \\
Annihilation of ${\mathrm{W^+ W^-}}$ bosons & <80.4 GeV $$$$ \\
		&& 76.5 & 86.25 & 86.25 & 86.25 \\
Annihilation of ${\mathrm{b \bar{b}}}$ quarks & <4190 MeV $$$$ \\
		&& 67.5 & 75.75 & 75.75 & 75.75 \\
Annihilation of ${\mathrm{\uptau^+ \uptau^-}}$ leptons & <1777 MeV $$$$ \\
		&& 64.5 & 72.25 & 72.25 & 72.25 \\
Annihilation of ${\mathrm{c \bar{c}}}$ quarks & <1290 MeV $$$$ \\
		&& 55.5 & 61.75 & 61.75 & 61.75 \\
QCD transition $^\dagger$ & 150--214 MeV$$$$ \\
		&& 15.5 & 17.25 & 17.25 & 17.25	\\
Annihilation of ${\mathrm{\uppi^+ \uppi^-}}$ mesons & <139.6 MeV $$$$ \\
		&& 13.5 & 15.25 & 15.25 & 15.25 \\
Annihilation of ${\mathrm{\uppi^0}}$ mesons & <135.0 MeV $$$$ \\
		&& 13.5 & 14.25 & 14.25 & 14.25 \\
Annihilation of ${\mathrm{\upmu^+ \upmu^-}}$ leptons & <105.7 MeV $$$$ \\
		&& 9.5 & 10.75 & 10.75 & 10.75 \\
Neutrino decoupling & <800 keV $$$$ \\
		&& 6.636 & 6.863 & 6.863 & 7.409 \\
Annihilation of ${\mathrm{e^+ e^-}}$ leptons & <511.0 keV $$$$ \\
		&& 3.636 & 3.363 & 3.363 & 3.909\\
\bottomrule
	\end{tabular}\\
	\begin{tabular}{lccccc}
\multicolumn{1}{c}{\footnotesize $^\dagger$ Using lattice QCD, this transition is normally calculated to 150--170 MeV.}
\end{tabular}
\end{table}

\subsection{Quark-Gluon Plasma~vs. Hadron Gas}
\label{Sec:QGPandHG}
In the early Universe, quarks and gluons moved freely around. A~gas consisting of quarks and gluons at high temperature is referred to as a~quark-gluon plasma, in analogy with an ordinary electromagnetic plasma. This is in contrast to today, where we do not observe free quarks, but only hadrons (e.g., pions and nucleons) that are bound states of either three quarks, three antiquarks, or a~quark-antiquark pair. These different combinations are called baryons and mesons, and are both bound together by the gluons. While quarks and gluons carry color charge, the hadrons we observe are color neutral. At some critical temperature of the Universe $T_c$, a~phase transition from a~quark-gluon plasma~to a~hadronic phase took place. We call the gas formed immediately after the phase transition a~hadron gas. This is similar to the formation of atoms, where the nucleus and electrons are bound together by electric forces. The aforementioned phase transition took place when the temperature of the Universe was approximately 150--170 MeV \cite{Petreczky:2012, Kapusta:QGP}. The transition temperature can be calculated by so-called lattice Monte Carlo simulations. Although the study of the phase transition from a~quark-gluon plasma~to a~hadron gas is rather difficult, we can get an estimate  of the critical temperature by evaluating the effective degrees of freedom for the energy density. This~estimate could be thought of as an upper limit bound, as we cannot have an increase in $\ge$ (i.e.,~the~energy density) as the universe expands

\subsection{Effective Degrees of Freedom in the QGP and HG Phases}
\label{Sec:dofQGPandHG}
\begin{figure}[H]
\centering
	\includegraphics[width=0.9\textwidth]{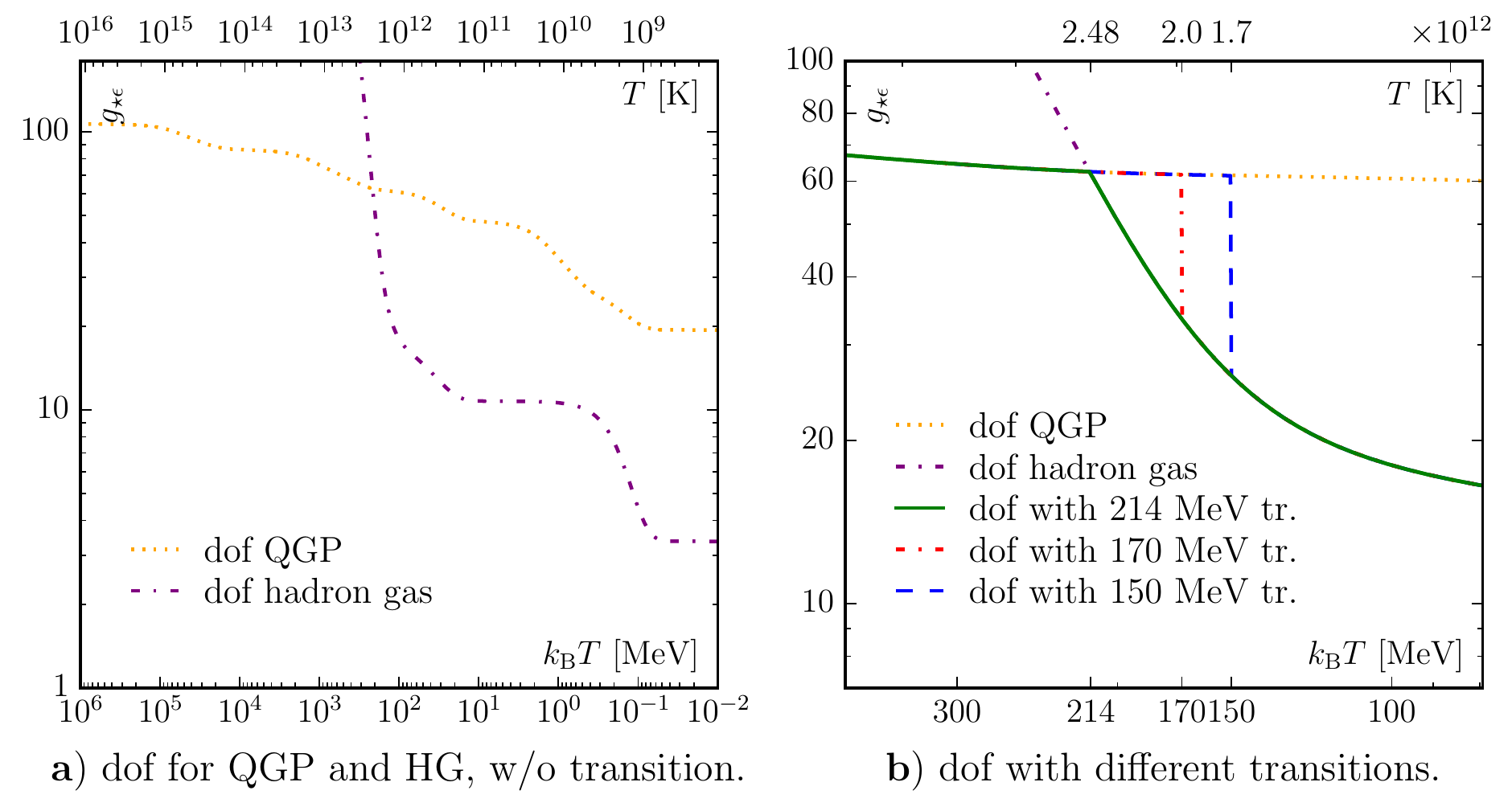}
	\caption{Panel (\textbf{a}) shows the effective degrees of freedom (dof) for $\ge$ in the quark-gluon and hadronic phases as functions of temperature. The yellow dotted curve represents the quark-gluon degrees of freedom and the purple dash-dotted curve is for the hadronic equivalent. In panel (\textbf{b}), we have zoomed in around the phase transition and plotted $\ge$ for three different transition temperatures: $\kB T_\textrm{c} = 214~\textrm{MeV}$ in solid green, $\kB T_\textrm{c} = 170~\textrm{MeV}$ in dash-dotted red, and $\kB T_\textrm{c} = 150~\textrm{MeV}$ in dashed blue.}
	\label{Fig:GStarComponents}
\end{figure}

Let us start an analysis at very high temperature, where all the elementary particles are present and effectively massless. $\ge$ is therefore at a~maximum. As the temperature decreases, the various particles annihilate, and $\ge$ falls accordingly. We trace the number of effective degrees of freedom as a~function of the temperature in Figure \ref{Fig:GStarComponents}a. Here, the yellow dotted curve shows the effective degrees of freedom in the quark-gluon plasma~phase (if it would exist for all temperatures). Without~a~phase transition, the quarks disappear only when the temperatures drop below their rest mass value. At the far right (colder) part of the scale, the gluons are still present together with the photons and neutrinos. In a~similar manner, we can trace the effective degrees of freedom in a~hadronic phase (if~it would exist for all temperatures), as shown in the purple dash-dotted curve. As in the real world, relatively speaking we only have photons and neutrinos present at low temperatures. As we go left to higher temperatures, the first increase in $\ge$ is caused by the presence of electron--positron pairs. The muons and the lightest mesons (namely, the pions), are the next particles to appear. We then get a~very steep increase in $\ge$, starting at around 100 MeV. This is due to the appearance of many heavier hadrons, whose~numbers grow almost exponentially as the temperature increases.

We can now define a~``cross-over'' temperature $T_\star$, which is the temperature at which the two curves intersect. Hence, the phase with the lower number of effective degrees of freedom for energy density wins (in QCD theory, one normally compares the pressure of the two phases, and the phase with the higher pressure wins). Using the particles listed by the Particle Data~Group \cite{PDG:2014} (and listed in Appendix \ref{Sec:MesonTable} and \ref{Sec:BaryonTable}), this yields $\kB T_\star=214$ MeV. However, if there are more possible baryonic states (which there most likely are), this temperature will be lower. This cross-over temperature could be thought of as the QCD transition temperature. To get a~more accurate estimate for the transition temperature, one can use the numerical method called lattice simulations. Using this latter method, one obtains a~transition temperature $\kB T_\textrm{c}$ in the 150--170 MeV range. The value depends on the number of quarks and their mass used for the calculation. Thus, our simple estimate gives us the correct order of magnitude, but a~bit too high. Speculatively, however, it is possible that
~it can be thought of as an upper bound.

In Figure \ref{Fig:GStarComponents}b, we zoom in around the transition temperature. We recognize the partly covered yellow and purple curves from panel-a, representing the QGP and HG scenarios. The green curve represents a~transition temperature of the aforementioned 214 MeV. If we insist on a~critical temperature of 170~MeV, we follow the yellow curve for the QGP to the right, and as we hit this temperature, we~jump down to the HG curve. This discontinuous curve for $\ge$ is shown in dash-dotted red color. We~will later see (Section \ref{Sec:Time}) that this can be interpreted as the temperature remaining constant over a~time while the degrees of freedom are reduced. The same remarks apply to the blue curve, which represents a~150 MeV transition.

\subsection{A~Closer Look at Each Particle Group}

\begin{figure}[H]
\centering
	\includegraphics[width=0.9\textwidth]{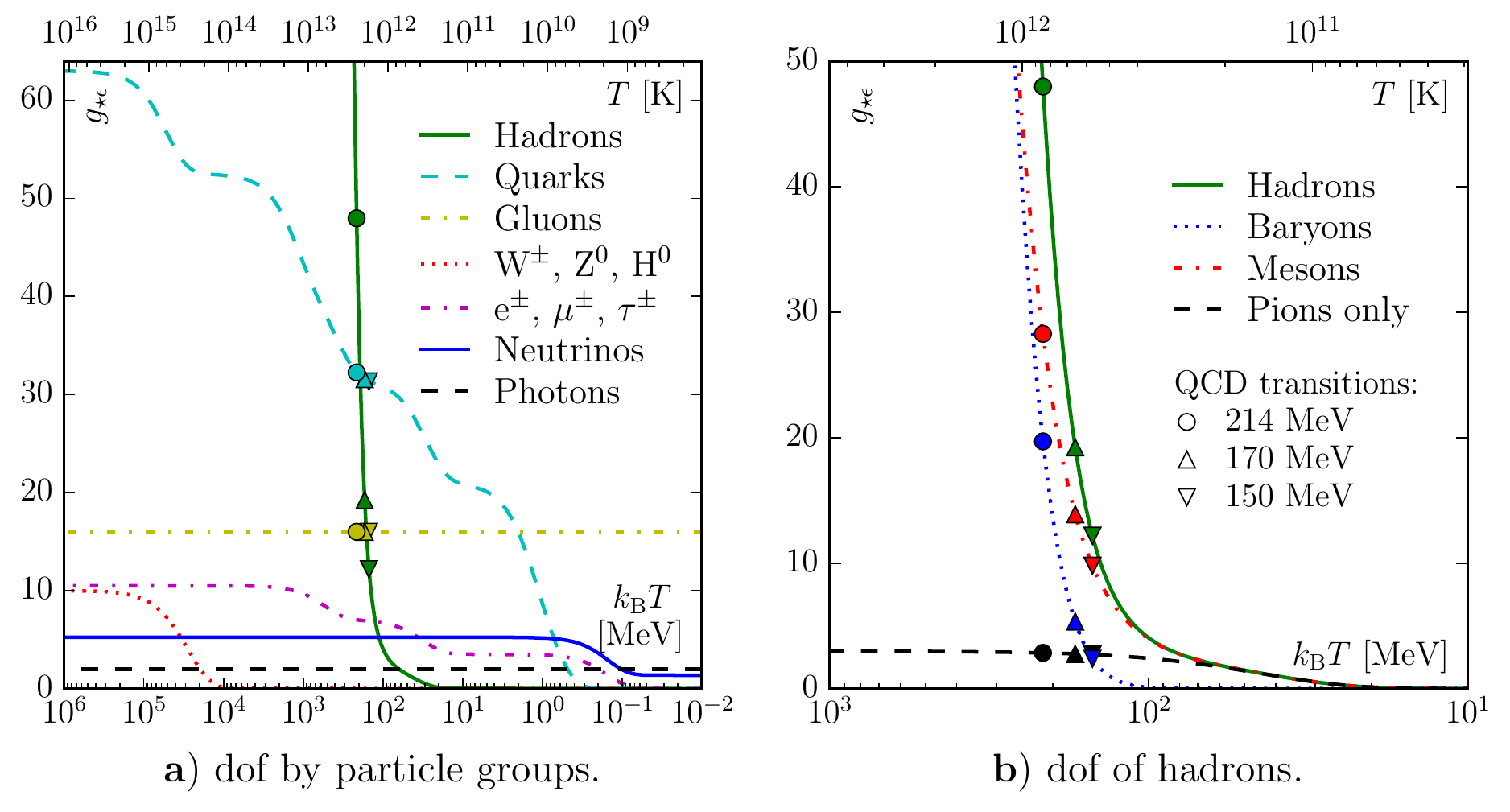}
	\caption{Panel (\textbf{a}) shows the contribution to the effective degrees of freedom (dof) for energy density from all particle groups. The drop in each group's $\ge$ value corresponds to ongoing annihilations of particles at that temperature. Panel (\textbf{b}) shows the total hadron contribution (green solid curve) to $\ge$ around the QCD phase transition temperature. We have further divided this into a baryon part (blue dotted curve) and a meson part (red dash-dotted curve). We have also plotted the pions specifically (black dashed curve), as they are the main hadronic contributor to $\ge$ at low temperatures. The two plots clearly show how fast the hadronic contribution increases at temperatures beyond 100 MeV. In both panels, we have marked the contribution to $\ge$ from hadrons, baryons, and mesons, at the three transition values of 214 MeV ($\medcircle$ symbols), 170 MeV ($\medtriangleup$ symbols), and 150 MeV ($\medtriangledown$ symbols), respectively.}
	\label{Fig:GStarSpecies}
\end{figure}

Let us have a~closer look at how each group of particle species contributes to $\ge$. Figure~\ref{Fig:GStarSpecies}a shows how the different particle groups contribute to the energy density as the temperature of the Universe drops. Let us look at the simplest case first---the photon (shown as the black dashed line). It~always has two degrees of freedom, and thus a~constant contribution, $\gee{\gamma}$, equal to two. The charged leptons (l) consist of the taus, muons, electrons, and their antiparticles. They are fermions, with two possible spin states. Each generation has a~degeneracy of 3.5, which adds up to 10.5 at high temperatures. The~magenta dash-dotted curve in Figure~\ref{Fig:GStarSpecies}a shows how the charged lepton contribution drops around the time when the temperature ($\kB T$) goes below that of the particle masses ($mc^2$).  The tau and antitau have a~mass of 1777~MeV, so when the temperature drops below this value, their abundance will drop, and at a~few hundred MeV they are all but gone, and $\gee{l}$ will have dropped to about 7. The~same process happens for the muons and electrons from $\kB T \sim 100~ \textrm{MeV}$ and  $\kB T \sim 0.5~ \textrm{MeV}$, when the value of $\gee{l}$ drops to 3.5, and finally zero. The case is more or less the same for the massive bosons (W$^\pm$, Z$^0$, and H$^0$). They have a~total degeneracy of 10, and all have masses of around 100 GeV, which means that their annihilations will overlap as seen in the red dotted curve. For neutrinos (solid blue curve), we see a~fall in $\gee{\nu}$ after they have decoupled, and the electron--positrons start to annihilate. We look closer at this in Section \ref{SubSec:NeutrinoDecoupling}.

For the color-charged particles (gluons and quarks), things are a~bit more complicated due to the differences before and after the QCD phase transition. In Figure \ref{Fig:GStarSpecies}a, we have plotted both the quark-gluon plasma~and hadron gas without any transition. Instead, we have marked their value at three different transition values: $\kB T_\textrm{c} = 214~\textrm{MeV}$ (marked~with~$\medcircle$), $\kB T_\textrm{c} = 170~\textrm{MeV}$ (marked~with~$\medtriangleup$), and $\kB T_\textrm{c} = 150~\textrm{MeV}$ (marked with $\medtriangledown$). The case for the gluons is straightforward---they have 16~degrees of freedom for $T > T_\textrm{c}$, and zero after. Quarks---being massive---begin with 63 effective degrees of freedom, which will gradually decrease as the top, bottom, and charm particles disappear. At the time of the phase transition, this value is down to about $\sim$32, depending on $T_\textrm{c}$.

After the phase transition, we need to count the hadronic degrees of freedom. We can distinguish these by baryons and mesons, as is done in Figure \ref{Fig:GStarSpecies}b. The only hadrons with masses less than $\kB T_\textrm{c}$ are the three pions, which for $T=T_\textrm{c}$ have roughly three degrees of freedom. There are, however, many heavier hadrons, which single-handedly do not contribute much at low temperatures, but the sheer number of different hadronic states results in a~collective significant contribution. Going from low to high temperatures in Figure \ref{Fig:GStarSpecies}b, the effective degrees of freedom from mesons (red dash-dotted curve) and baryons (blue dotted curve) increase almost exponentially. This value is quite different at different $T_\textrm{c}$. Following the hadrons (green curve) from right to left, we see that at $\kB T=150~\textrm{MeV}$, the hadrons make up roughly 12 effective degrees of freedom. At $\kB T=170~\textrm{MeV}$, this number is approximately $19$, and at $\kB T=214~\textrm{MeV}$, we have roughly $48$---which is the same as the $16+32$ degrees of freedom from the free quarks and gluons.

\section{Decoupling}
\label{Sec:Decoupling}
As we mentioned in Section \ref{Sec:Introduction}, particles are kept in thermal equilibrium by constantly colliding (interacting) with each other. The collision rate depends on two factors---the cross section $\sigma$ and the particle density $n$. The cross section depends on several factors, but the most important one is by~which forces the particles interact. Those which feel the strong and electromagnetic force interact strongly, while those which only feel the weak force interact much weaker. The cross sections related to the different forces depend on the temperature, or more correctly on the energy involved in the reaction. How these interaction strengths change are different for the four forces. In general, they become closer in strength for higher temperatures.

When the Universe expands, dilutes, and cools, particles travel farther and farther before interacting. That is, their mean free path and lifetime increases. As mentioned in Section \ref{Sec:Evolution}, at some time the interaction rate for some particles can become slower than the expansion rate of the Universe, and (on average) those particles will never interact again. The time at which this happens is defined as the time of decoupling. For neutrinos, this happened about one second after the Big Bang (and we will get back to this in the next section), and for photons this happened about $380,000$ years later (due to recombination and forming of neutral atoms). Let us look at the general case. First we need to introduce the concept of comoving coordinates and volumes. Comoving coordinates move with the rest frame of the Universe; i.e., they do not change as the Universe expands. An analogy of this would be to draw dots on a~balloon. The actual distance between the dots increases as the balloon is inflated, but their comoving distance remains the same. For a~comoving volume with constant entropy $S$, we can write
\begin{align}
  S = s(T) a^3 = \gs(T) \frac{2\pi^2}{45} T^3 a^3 &= \textrm{constant} \notag \\
  \rightarrow \quad  \gs(T) T^3 a^3 &= \textrm{constant} \;.
  \label{Eq:gTa}
  \end{align}

One of the consequences of this is that the temperature will fall slower during particle annihilations (i.e., when the effective degrees of freedom decreases). To understand this, we need to look at what is going on during particle creations and annihilations, as well as rest mass energy vs. kinetic energy.

During reactions where we go from two massive particles to two lighter particles, the excess rest mass energy will be converted to kinetic energy. Thus, the lighter particles will on average have a~higher kinetic energy than the other particles in the thermal ``soup''. Normally, this is countered by the reversed reaction---namely, reactions where two lighter particles create two more massive ones with less kinetic energy.
Throughout periods where we have particle annihilations, there will be a~net flow of massive particles to lighter particles plus kinetic energy. Hence, the temperature will fall slower in these periods.

In order to maintain thermal equilibrium, particles need to constantly interact. That~is, there~needs to be some coupling between them (directly or indirectly). If some particles decouple, it~means that they on average will never interact again, so if a~particle species has decoupled before an annihilation process starts, their temperature will decrease independently of those which are still coupled together. As~a~result, there will be two different temperatures: the photon-coupled temperature ($T$) (those particles that directly or indirectly interact with the photons), and the decoupled-particle temperature ($T_\textrm{dc}$). Solving~Equation~(\ref{Eq:gTa}) before  and after an annihilation process (indicated by subscripts ``1'' and ``2'') for the photon-coupled ($\gamma \mathrm{c}$) and decoupled ($\mathrm{dc}$) particles gives us
\begin{equation}
  g_{\gamma\textrm{c}1} T_1^3 a_1^3 = g_{\gamma\textrm{c}2} T_2^3 a_2^3 \;,
  \label{Eq:ggc}
  \end{equation}
\begin{equation}
  g_{\textrm{dc}1} T_{\textrm{dc}1}^3 a_1^3 = g_{\textrm{dc}2} T_{\textrm{dc}2}^3 a_2^3 \;.
  \label{Eq:gdc}
  \end{equation}

After decoupling, but before an annihilation process, the two temperatures are the same. Well, close enough, as we will briefly discuss in Section \ref{Sec:Time}. Once a photon-coupled particle species start to annihilate, the~degrees of freedom for (all) the coupled particles will reduce, while it will remain the same for the decoupled ones. Solving for the decoupled temperature after annihilation gives us:
\begin{equation}
  T_{\textrm{dc}2}^3 = \frac{g_{\gamma\textrm{c}2}}{g_{\gamma\textrm{c}1}} T_2^3 \;,
  \label{Eq:Tdc3}
  \end{equation}
which we normally write as
\begin{equation}
  T_{\textrm{dc}} = \sqrt[3]{\frac{g_{\gamma\textrm{c}2}}{g_{\gamma\textrm{c}1}}} T \;.
  \label{Eq:Tdc}
  \end{equation}

In principle, we can do this for more than one decoupled particle species, and get two or more different temperatures for the decoupled particles.

\subsection{Neutrino Decoupling}
\label{SubSec:NeutrinoDecoupling}
Before they are decoupled, neutrinos are kept in thermal equilibrium with the photon-coupled particles mainly via~weak interactions with electrons and positrons. Around one second after the Big Bang, the rate of the neutrino--electron interactions becomes slower than the rate of expansion of the Universe, $H$. The collision rate between neutrinos and electrons (and its antiparticle), $\Gamma_{\nu}$, is given by
~\cite{Weinberg:Cosmology, Lesgourgues:2012}:
\begin{align}
  \Gamma_{\nu}	= 		 n_\textrm{e} \sigma_{\textrm{wk}}
  				&\approx \left( \frac{\kB T}{\hbar c} \right)^3 (\hbar c \Gw \kB T)^2 \notag \\
			    &\approx \frac{\Gw^2(\kB T)^5}{\hbar c} \;,
  \label{Eq:NeutrinoElectronCollisionRate}
  \end{align}
where $n_e$ is the number density of electrons and $\sigma_\textrm{wk}$ is the neutrino--electron scattering cross section. $\Gw = \GF/(\hbar c)^3 \approx 1.166 \times 10^{-5}$ GeV$^{-2}$ is the weak coupling constant \cite{Griffiths:Particles, PDG:2012}. By using the equation for energy density, either from Equation (\ref{Eq:ej}c), or better, by fast-forwarding to Equation (\ref{Eq:EnergyDensityByG}), the expansion rate at the same time is given by the first Friedmann equation:
\vspace{12pt}
\begin{align}
  H = \sqrt{\frac{8\pi \GN}{3c^2} \ee}
    &= \sqrt{\frac{8\pi \GN}{3c^2} \ge(T) \frac{\pi^2}{30}\frac{(\kB T)^4}{(\hbar c)^3}} \notag \\
    &\approx \sqrt{\frac{5 \GN(\kB T)^4}{(\hbar c)^3}}  \;.
\end{align}

The prefactors in $\Gamma_\nu$ and $H$ roughly cancel
each other, such that we end up with
\begin{equation}
  \frac{\Gamma_{\nu}}{H} \approx \Gw^2 \sqrt{\frac{\hbar c}{5\GN}}(\kB T)^3
  \approx \left(\frac{T}{10^{10} \textrm{ K}} \right)^3 \;.
  \label{Eq:NeutrinoDecouplingTemperature}
  \end{equation}

This is a~rough estimate, but one that is most commonly used (e.g., by Weinberg \cite{Weinberg:Cosmology}). Being~relativistic, the~neutrino temperature $T_{\nu}$ scales as $a^{-1}$, while the energy density and number density scale as $a^{-4}$ and $a^{-3}$, respectively.

\subsection{Neutrino Temperature and Entropic Degrees of Freedom}
\label{SubSec:NeutrinoTemperature}
Let us look more closely at the effective degrees of freedom at the time just after the neutrinos decouple. For the entropy density before the electrons and positrons annihilate, they have 10.75 degrees of freedom, divided as $5.25$ for the neutrinos and $2+3.5=5.5$ for the photon plus the electron and positron. The latter one is reduced to just $2$ once all the electrons and positrons have annihilated (i.e.,~$g_{\gamma\textrm{c2}}/g_{\gamma\textrm{c1}}=2/5.5=4/11$). We now have a~higher photon temperature and a~lower neutrino temperature. Using Equation (\ref{Eq:Tdc}), we find the neutrino temperature after all electrons and positrons have annihilated to~be
\begin{equation}
  T_{\nu}=\sqrt[3]{\frac{4}{11}} T \simeq 0.71 T \;.
  \label{TNu}
  \end{equation}

Hence, after the electron--positron annihilation, the neutrino temperature is 71\% that of the photon temperature. Measurements of the Cosmic Microwave Background (CMB) radiation is found to be 2.73~K. This means that the neutrino background temperature should be 1.95 K (it should be mentioned that no measurement of the cosmic neutrino background have been made, or is likely to be made in the near future that would confirm this prediction).

The colder neutrinos do not contribute as much as the hotter particles to the four different $\g$s, and this has to be taken into account when we calculate the different effective degrees of freedom. In~general, after a~particle species decouples, we need to introduce a~species-dependent temperature ratio into our equations; that is,  $T \rightarrow T(T_j/T)$. Here $T_j$ is the temperature of the decoupled particle species, while $T$ is the photon-coupled (reference) temperature. We thus get the following $\gn$, $\ge$, $\gp$, and~$\gs$
after electron--positron annihilation is completed
\begin{align}
  \gn 		&= 2 + 6 \times \frac{3}{4} \left( \frac{T_\nu}{T} \right)^3
			 = 2 + 6 \times \frac{3}{4} \times \frac{4}{11}
			 = \frac{40}{11} \approx 3.636 \;. \\
  \ge=\gp	&= 2 + 6 \times \frac{7}{8} \left( \frac{T_\nu}{T} \right)^4
			 = 2 + 6 \times \frac{7}{8} \left( \frac{4}{11} \right)^{4/3}
			 \approx  3.363 \;, \\
  \gs 		&= 2 + 6 \times \frac{7}{8} \left( \frac{T_\nu}{T} \right)^3
			 = 2 + 6 \times \frac{7}{8} \times \frac{4}{11}
			 = \frac{43}{11} \approx 3.909 \;.
  \end{align}

 The neutrino contribution during electron--positron annihilation is found by subtracting the electron--positron contribution in the following way:

\begin{align}
  \gnn{_\nu} &= \frac{6 \times 3}{4} \times \left[ \frac{4}{11} +
\left(1-\frac{4}{11} \right) \frac{4}{4 \times 3} \gnn{_\mathrm{e}} \right] \;,
\end{align}
\vspace{12pt}
\begin{align}
  \gee{_\nu} 	&= \frac{6 \times 7}{8} \times \left[
\left(\frac{4}{11}\right)^{4/3} + \left(1-\left(\frac{4}{11}\right)^{4/3} \right)
\frac{8}{4 \times 7}  \gee{_\mathrm{e}} \right] \;, \\
  \gpp{_\nu} 	&= \frac{6 \times 7}{8} \times \left[
\left(\frac{4}{11}\right)^{4/3} + \left(1-\left(\frac{4}{11}\right)^{4/3} \right)
\frac{8}{4 \times 7}  \gpp{_\mathrm{e}} \right] \;, \\
  \gss{_\nu} 	&= \frac{6 \times 7}{8} \times \left[ \frac{4}{11} +
\left(1-\frac{4}{11} \right) \frac{8}{4 \times 7} \gss{_\mathrm{e}} \right] \;,
\end{align}
where the four $g_{\star x_\mathrm{e}}$ are the effective electron--positron contributions.

In reality, as can be seen in Figure \ref{Fig:BosonFermionIntegrals} and Table \ref{Tab:OneParticleContribution}, the first electron--positron annihilations began slightly before the neutrino decoupling was complete. Hence, some of the energy from the decaying electron--positron pairs heated up the neutrinos. This caused a~small deviation from the above-mentioned values, which resulted in effective numbers of neutrino species slightly larger than three. This number is given to be 3.046 by Mangano \cite{Mangano:2005} and 3.045 by de Salas and Pastor~\cite{deSalas:2016}. By using Mangano's result, a~compensated result will be
\begin{align}
  \gn &= 	2 + 2 \times 3.046 \times \frac{3}{4} \times \frac{4}{11}
  \approx 3.661 \;, \\
  \ge = \gp &= 2+2\times 3.046 \times \frac{7}{8} \left(\frac{4}{11}
  \right)^{4/3}\approx 3.384 \;,\\
  \gs &= 2 + 2 \times 3.046  \times \frac{7}{8} \times \frac{4}{11}
  \approx 3.938 \;.
  \end{align}

\section{Functions for \boldmath{$\nn$, $\ee$, $\pp$}, and \boldmath{$S$}, and Their Implications}
We can now express the complete number density, energy density, pressure, and entropy density in terms of their effective degrees of freedom:
\begin{empheq}{align}
  \textrm{Number density:} \quad \;\;\;\,
  \nn(T) &= \frac{\zeta(3)}{\pi^2} \gn(T) \frac{(\kB T)^3}{(\hbar c)^3} \;,
  \label{Eq:NumberDensityByG} \\
  \textrm{Energy density:}  \qquad\;\;
  \ee(T) &= \frac{\pi^2}{30} \ge(T)  \frac{(\kB T)^4}{(\hbar c)^3} \;,
  \label{Eq:EnergyDensityByG}\\
  \textrm{Pressure:} \qquad \qquad \quad
  \pp(T) &= \frac{\pi^2}{90} \gp(T)  \frac{(\kB T)^4}{(\hbar c)^3} \;,
  \label{Eq:PressureByG}\\
  \textrm{Entropy density:} \qquad
  \ss(T) &= \frac{2\pi^2}{45} \gs(T)  \frac{\kB^{\;4} T^3}{(\hbar c)^3} \;.
  \label{Eq:EntropyByG}
\end{empheq}

\begin{figure}[H]
\centering
	\includegraphics[width=0.88\textwidth]{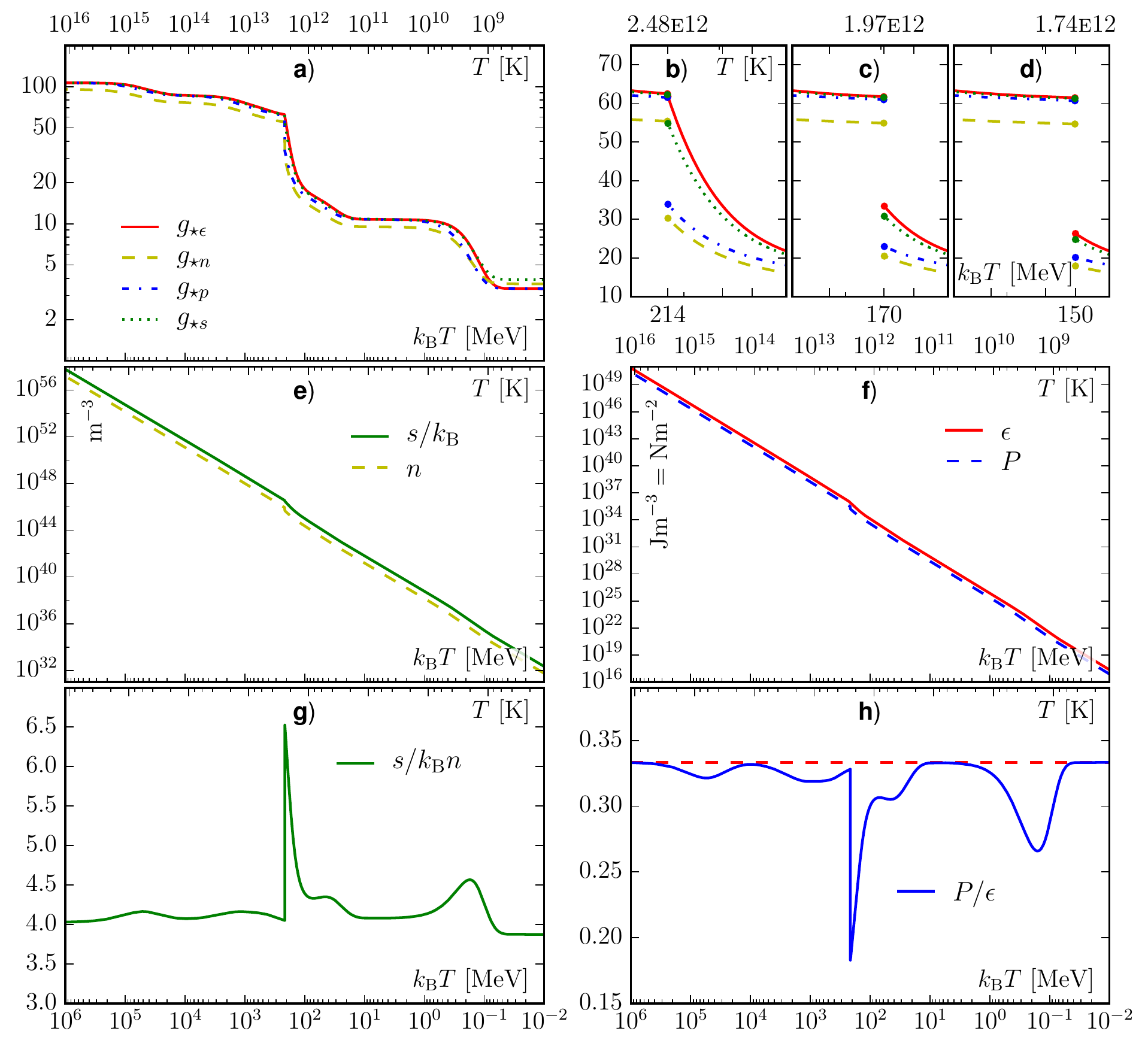}
	\caption{Panel (\textbf{a}) shows the four $\g$s. At $\kB T = 214$ MeV, only $\ge$ is continuous, while~$\gn$, $\gp$, $\gs$ drops in value by between 10 and 30. The three upper right panels show these jumps at transition values of 214 MeV (\textbf{b}), 170 MeV (\textbf{c}), and 150 MeV (\textbf{d}). Panels (\textbf{e}) and (\textbf{f}) show the evolution of $\nn$, $\ee$, $\pp$, and $\ss/\kB$ as a function of temperature. In the two lower panels, we look at the relation between entropy density and number density (\textbf{g}) and pressure and energy density (\textbf{h}). We see small fluctuations during periods with particle annihilations, especially right after the QCD phase transition. As we remember from Figure \ref{Fig:BosonFermionIntegrals}, this is because the pressure and number density drop quicker than energy density and entropy density at these times. The short physical explanation is that semi- and non-relativistic particles exert less pressure and have a~higher entropy than relativistic particles (at the same temperature). One consequence of this is that particle numbers are not conserved, which is not a~requirement for particles whose chemical potential is zero.}
	\label{Fig:NEPS}
\end{figure}

We have plotted these functions as well as the $\g$ values in Figure \ref{Fig:NEPS}. The energy density and pressure have the same dimension, while the dimensions of entropy density and number density differ by the Boltzmann constant (unit: J K$^{-1}$).

When the prefactors are accounted for, the difference in $\ss$ and $\nn$, and $\pp$ and $\ee$, lies in the deviations between $\gs$ and $\gn$, and $\gp$, and $\ge$. So, let us discuss a~bit more about what is actually happening. Both the increase in entropy per particle and the decrease in pressure (which we see as bumps and dips in panels (\textbf{g}) and (\textbf{h}) in Figure \ref{Fig:NEPS}) are due to the presence of particles at semi- and non-relativistic temperatures. Before we go any farther, we should address the QCD phase transition. As not all four $\g$s can be continuous (as we see in panels (\textbf{b})--(\textbf{d}) in Figure \ref{Fig:NEPS}), we get inconsistencies and some unphysical results. For most of our plots, we use $T_\textrm{c} = 214~\textrm{MeV}$, keeping $\ge$ continuous, leaving $\gn$, $\gp$, and $\gp$ discontinuous at this point.


We see from panel (\textbf{a}) in Figure \ref{Fig:NEPS} that both number density and entropy density decrease more rapidly during annihilation periods. However, this is a~bit deceiving, since we are looking at their values as functions of temperature. In fact, the total entropy stays constant (it actually~increases ever so slightly if we do not have perfect thermal equilibrium). 
Both $\ss$ and $\nn$ fall a~bit as~we cross the transition temperature. We thus get a~jump in the entropy per particle at this time, as~can be seen in panel (\textbf{g}) in Figure \ref{Fig:NEPS}. The other bumps in entropy per particle are continuous. The entropy per particle will start to rise when the rest mass of some massive particles becomes more significant. $s$ then flattens out and drops again as these particles gradually become less numerous. After all these massive particles have annihilated and disappeared, the value of $s$ returns to its original value (before the annihilations started). As~mentioned in Figure \ref{Fig:BosonFermionIntegrals} in Section \ref{Sec:EffectiveDegreesOfFreedom}, particles whose rest mass energy is significant have a~higher total energy, and thus a~higher entropy. This entropy is eventually transferred to the remaining particles after they annihilate. It is important to emphasize that it is not the total entropy that changes, but~rather the (total) particle number that falls and rises again. The change in entropy density after the neutrino decoupling, as we can see at the lowest temperature in panel (\textbf{b}) in Figure \ref{Fig:NEPS}, is due to the fact that the neutrinos have a~lower temperature and thus contribute less.

The same explanation goes for the fall in pressure, as seen in panel (\textbf{h}) in Figure \ref{Fig:NEPS}. Non-relativistic particles exert (relatively) zero pressure. The pressure is thus at its lowest at times where the ratio of semi- and non-relativistic particles are at their highest. We see that the two most significant drops in pressure are just after the QCD phase transition and in the middle of the electron--positron annihilations. We have used a~naive definition for our QCD phase transition---namely, that of the lowest energy density. In reality, this transition is quite complex, and we should interpret our result with a~grain of salt. With that in mind, we go from an almost pure relativistic gas (QGP) to a~case where the majority of the particles are semi- or non-relativistic (HG)---which is the reason for the jump down in pressure at $T=214$ MeV.


As we will get back to in the next section, the Universe expands faster when it is matter-dominated as compared to when it is radiation-dominated. So even though the early Universe was the latter, we know from our study that we have periods with a~significant fraction of semi-relativistic particles. One can thus argue that $a$ should grow slightly faster at these times.

\section{Time--Temperature Relation}
\label{Sec:Time}

As mentioned in Section \ref{Sec:Introduction}, the measurements of the CMB thermal spectrum is very close to that of a~perfect black body \cite{Fixsen:2009}. The early Universe should be very homogeneous, with the same features everywhere.
How fast the early Universe expands depends on which energy contributor is dominating---the relativistic particles (radiation), or non-relativistic particles (cold matter).
By~solving the Friedmann equations for a~flat adiabatic Universe with no cosmological constant, we find the relation between the scale factor ($a$) and time ($t$) to be $a=t^{1/2}$ for a~pure radiation case, and $a=t^{2/3}$ for a~pure cold matter case. Simple derivations for this are given by Ryden \cite{Ryden:IntroductionToCosmology} and Liddle \cite{Liddle:AnIntroductionToModernCosmology}. Similarly,~a~relation between the scale factor and temperature for the two extreme cases is given as $T=a^{-1}$ and $T=a^{-2}$ for the two cases \cite{Kolb:EarlyUniverse, Rahvar:2006}. This gives us the following relation between the three~quantities:
\begin{align}
  \textrm{Just radiation:} \qquad \quad
	&T \propto t^{-1/2} 	\propto a^{-1} 	\;, \\
  \textrm{Just cold matter:} \qquad
	&T \propto t^{-4/3} 	\propto a^{-2} 	\;.
  \label{Eq:TimeTemperatureScale}
  \end{align}

For a~mixture of both types of particles, we should have something in between the two single-component cases. So, if radiation is the more dominant energy contributor, $a$ grows almost proportional to $t^{1/2}$, or more proportional $t^{2/3}$ for the matter case.
Regarding temperature, for a~radiation-dominated scenario, the relation between temperature and time (after the Big Bang) can be calculated as a~function of $\ge$, as follows \cite{PDG:2014}:
\begin{align}
  t &= \sqrt{\frac{90 \hbar^3 c^5}{32 \pi^3 \GN \ge(T)}} (\kB T)^{-2} \notag\\
	&= \frac{2.4}{\sqrt{\ge(T)}} T^{-2}_{\textrm{MeV}} \;.
  \label{Eq:TimeTemperatureRadiationEra}
  \end{align}

The Universe becomes matter-dominated at roughly $10^5$ years after the Big Bang, long after the scope of this article. However, it should be noted that the temperature of both photons and matter drop as the inverse of the scale factor, even long after this radiation--matter equality. This is because temperature is determined by the kinetic energy of the particles, while the definition of a~radiation- or matter-dominated Universe is that of the total energy (where rest mass is included). Being outnumbered more than a~billion to one, matter is unable to cool down the photons, and the temperature of the Universe drops as the inverse of the scale factor until matter decouples from the photons.



\begin{figure}[H]
	\centering
	\includegraphics[width=0.7\textwidth]{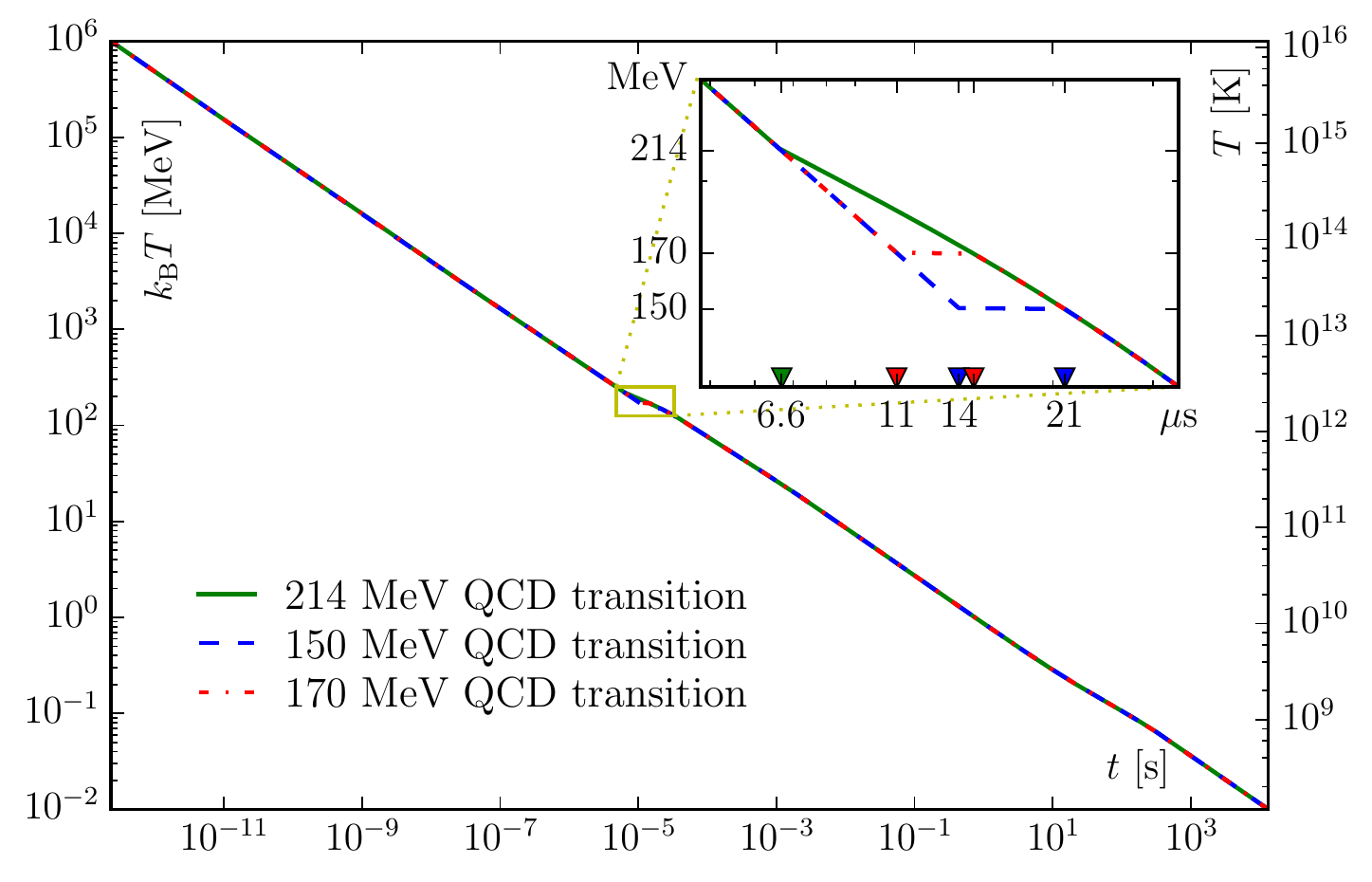}
	\caption{Energy and temperature as functions of time using the three different transition temperatures. For the $\kB T_\textrm{c} = 214~\textrm{MeV}$ transition, the temperature will drop slower after $T_\textrm{c}$, but nonetheless always decrease over time. For $\kB T_\textrm{c} = 170~\textrm{MeV}$ and $\kB T_\textrm{c} = 150~\textrm{MeV}$, there will be a~period with constant temperature and energy density while $\ge$ decreases from its quark-gluon value to its hadron gas value.}
	\label{Fig:Time}
\end{figure}

On the other hand, when massive particles die out, their annihilation energy is transferred to the remaining particles in the thermal bath. This should make the temperature drop slower as a function of time. If we assume that this latter argument is dominant, we will get a~Universe which drops  in temperature more slowly when $\ge$ is decreasing. Figure \ref{Fig:Time} shows the temperature as a function of time assuming a~pure radiation-dominated Universe, as given by Equation (\ref{Eq:TimeTemperatureRadiationEra}). During particle annihilations, we have a~smooth continuous function, but this is not the case for the QGP-to-HG transition using our models. Here we have to consider  the three different transition temperatures separately. Using $\kB T_\textrm{c} = 214~\textrm{MeV}$, we have a~scenario where the temperature will drop more slowly right after $T_\textrm{c}$. Using $\kB T_\textrm{c} = 170~\textrm{MeV}$ and $\kB T_\textrm{c} = 150~\textrm{MeV}$, the degrees of freedom ($\ge$) will jump down at $T_\textrm{c}$. For $\kB T_\textrm{c} = 170~\textrm{MeV}$, the value of $\ge$ falls from around 62 to 33, while for $\kB T_\textrm{c} = 150~\textrm{MeV}$, this value falls from around 61 to 26. This~would, however, take some time. Using our simple model, the temperature and energy density of the Universe would stay constant as the Universe expands until it reaches its pure hadron-gas state.

The relation made here between time and temperature during the QCD transition is naive and simple, and the numerical values thereafter. Small deviations from our plot during the QCD transition (and for that matter, during regular particle annihilations) only affect the period on hand, and become negligible as time go on.

\section{On the QCD Phase Transition and Cross-Over Temperature}
Our method of using $\kB T_\textrm{c} = 214~\textrm{MeV}$ is based on a~calculation where we add all the $\ge$ from the quark-gluon state on one hand (easy), and all the $\ge$ from the hadrons on the other hand (not so easy). We have used the hadronic particles as listed in Appendix \ref{Sec:MesonTable} and \ref{Sec:BaryonTable}. These are the particles listed by the \textit{Particle Data~Group} \cite{PDG:2014}. There are additional candidates to these lists---some good candidates, and some more speculative. There could also be more hadronic states which are hard to detect. For every new hadron we add to our model, the cross-over temperature decreases. Not so much for the most massive candidates, but more so for the lighter ones.

The phase transition we have used is of first order. Only $\ge$ is continuous at this point, and $\gn$, $\gp$, and $\gs$ are not. This will lead to some unphysical consequences---such as an instantaneous increase of the scale factor by roughly 4\% if we assume constant entropy. A~proper theory about the QCD phase transition is required to address this issue, which is beyond the scope of this article. The theory we have presented here should be valid to a~good approximation, before and after the QCD~transition.

\section{Conclusions}
Our knowledge of the very first stages of Universe is limited. In order to know for sure what is happening at these extreme energies and temperatures, we want to recreate the conditions using particle accelerators. The Large Hadron Collider at CERN can collide protons together at energies of 13 TeV, and with their discovery of the Higgs boson, all the elementary particles predicted by the Standard Model of particle physics have been found. This is, however, most likely not the complete story. Dark~matter particles are the hottest candidates to be added to our list of particles, and there is almost sure to be more particles at even higher temperatures, such as at the Grand Unified Theory (GUT) scale of $T \sim 10^{16}$ GeV.

We have here used the statistical physics approach to counting the effective degrees of freedom in the early Universe at temperatures below 10 TeV. Some simplifications have been used, such as setting the chemical potential equal to zero for all particles. The aim of this article was to give a~good qualitative introduction to the subject, as well as providing some quantitative data~in the form of plots and tables. 

The early Universe is often thought of as being pure radiation (just relativistic particles). However,~when the temperature drops to approximately that of the rest mass of some massive particles, we~get interesting results, where we have a~mix of relativistic particles and semi- and non-relativistic ones. This mix is most prominent during the electron--positron annihilations, and just after the phase transition from a~quark-gluon plasma~to a~hadron gas. Approaching this, using our ``no-chemical potential'' distribution functions shows us how the entropy per particle increases when the ratio of semi- and non-relativistic particles becomes significant.

The number of effective degrees of freedom for hadrons changes very quickly around the QCD transition temperature. We found a~cross-over temperature of 214 MeV using the known baryons and mesons. As there could be many more possible hadronic states than we have accounted for, this cross-over temperature could be lower. This is not meant as a~claim of a~new QCD transition temperature, but rather as an interesting fact. Our first-order approach based purely on the distribution functions has inconsistencies at the cross-over temperature.

We have listed the effective degrees of freedom for number density ($\gn$), energy density ($\ge$), pressure ($\gp$), entropy density ($\gs$), and time ($t$) as function of temperature ($T$) in Table \ref{Tab:Values} in Appendix~{\ref{Sec:GTable}. Table \ref{Tab:OneParticleContribution} in Appendix \ref{Sec:OneParticleContribution} lists the different effective contributions to a~single intrinsic degree of freedom, corresponding to our plot in Figure \ref{Fig:BosonFermionIntegrals}.

\appendixtitles{yes}
\appendixsections{multiple}
\appendix
\section{Table for Time, $\gn$, $\ge$, $\gp$, and $\gs$} \label{Sec:GTable} 
\makeatletter
\setcounter{table}{0}
\@addtoreset{table}{section}
\renewcommand{\thetable}{A\arabic{table}}
\makeatletter
\vspace{-16pt}
\begin{footnotesize}
\begin{longtable}{ccccccccc}
\caption{Values for time, $\gn$, $\ge$, $\gp$, and $\gs$ from $T=10~\textrm{TeV}$ to $10~\textrm{keV}$. In the region between 150--214 MeV, the values for all five quantities depend on the model's transition temperature. We should emphasize that the values for this period are based on our simple no-chemical potential model.} \label{Tab:Values} \\
\toprule
$\bm{\kB T}$ \textbf{(eV)}
& $\bm{T}$ \textbf{(K)}
& \textbf{Time ($\bm{s}$)}
& $\bm{\gn}$ \phantom{.}
& $\bm{\ge}$ \phantom{.}
& $\bm{\gp}$ \phantom{.}
& $\bm{\gs}$ \phantom{.}
& \textbf{Tr. Temp.} \\
\midrule \endfirsthead
\caption{{\em Cont.}} \label{Tab:Values} \\
\toprule
$\bm{\kB T}$ \textbf{(eV)}
& $\bm{T}$ \textbf{(K)}
& \textbf{Time ($\bm{s}$)}
& $\bm{\gn}$ \phantom{.}
& $\bm{\ge}$ \phantom{.}
& $\bm{\gp}$ \phantom{.}
& $\bm{\gs}$ \phantom{.}
& \textbf{Tr. Temp.} \\
\midrule \endhead
\bottomrule \endfoot
10 TeV	& 1.16\E{17}  & 2.32\E{-15} &  95.50 & 106.75 & 106.75 & 106.75 & All \\
  5 TeV	& 5.80\E{16}  & 9.29\E{-15} &  95.49 & 106.75 & 106.75 & 106.75 & All \\
  2 TeV	& 2.32\E{16}  & 5.81\E{-14} &  95.47 & 106.74 & 106.73 & 106.74 & All \\
  1 TeV	& 1.16\E{16}  & 2.32\E{-13} &  95.39 & 106.72 & 106.65 & 106.70 & All \\
500 GeV	& 5.80\E{15}  & 9.30\E{-13} &  95.11 & 106.61 & 106.38 & 106.56 & All \\
200 GeV	& 2.32\E{15}  & 5.83\E{-12} &  93.55 & 105.90 & 104.75 & 105.61 & All \\
100 GeV	& 1.16\E{15}  & 2.36\E{-11} &  89.89 & 103.53 & 100.80 & 102.85 & All \\
 50 GeV	& 5.80\E{14}  & 9.73\E{-11} &  83.53 &  97.40 &  93.94 &  96.53 & All \\
 20 GeV	& 2.32\E{14}  & 6.40\E{-10} &  77.39 &  88.45 &  87.22 &  88.14 & All \\
 10 GeV	& 1.16\E{14}  & 2.58\E{-9}  &  76.20 &  86.22 &  85.85 &  86.13 & All \\
  5 GeV	& 5.80\E{13}  & 1.04\E{-8}  &  75.27 &  85.60 &  84.68 &  85.37 & All \\
  2 GeV	& 2.32\E{13}  & 6.61\E{-8}  &  71.14 &  82.50 &  79.69 &  81.80 & All \\
  1 GeV	& 1.16\E{13}  & 2.75\E{-7}  &  65.37 &  76.34 &  72.97 &  75.50 & All \\
500 MeV	& 5.80\E{12}  & 1.15\E{-6}  &  59.69 &  69.26 &  66.43 &  68.55 & All \\				
214$^+$ MeV&2.48\E{12}& 6.63\E{-6}	&  55.37 &  62.49 &  61.52 &  62.25 & All \\
214$^-$ MeV&2.48\E{12}& 6.63\E{-6}	&  30.27 &  62.49 &  33.88 &  54.80 & 214 MeV  \\
		& 			  & 6.63\E{-6}	&  55.37 &  62.49 &  61.52 &  62.25 & 170 $+$ 150 MeV \\
200 MeV	& 2.32\E{12}  & 8.42\E{-6}	&  26.45 &  50.75 &  29.62 &  45.47 & 214 MeV \\
		&		 	  & 7.61\E{-6}	&  55.21 &  62.21 &  61.34 &  61.99 & 170 $+$ 150 MeV \\
190 MeV	& 2.20\E{12}  & 1.00\E{-5}	&  24.14 &  44.01 &  27.04 &  39.77 & 214 MeV \\
		&			  & 8.44\E{-6}	&  55.10 &  62.03 &  61.21 &  61.83 & 170 $+$ 150 MeV \\
180 MeV	& 2.09\E{12}  & 1.20\E{-5}	&  22.16 &  38.27 &  24.84 &  34.91 & 214 MeV \\
		&			  & 9.42\E{-6}	&  54.99 &  61.87 &  61.07 &  61.67 & 170 $+$ 150 MeV \\
170$^+$ MeV&1.97\E{12}& 1.44\E{-5}	&  20.49 &  33.47 &  22.98 &  30.84 & 214 MeV \\
		&			  & 1.06\E{-5}	&  54.88 &  61.72 &  60.94 &  61.52 & 170 $+$ 150 MeV \\
170$^-$ MeV&1.97\E{12}& 1.44\E{-5}	&  20.49 &  33.47 &  22.98 &  30.84 & 214 $+$ 170 MeV \\
		&			  & 1.06\E{-5}	&  54.88 &  61.72 &  60.94 &  61.52 & 150 MeV \\
160 MeV	& 1.86\E{12}  & 1.73\E{-5}	&  19.09 &  29.51 &  21.42 &  27.49 & 214 $+$ 170 MeV \\
		&			  & 1.20\E{-5}	&  54.77 &  61.58 &  60.80 &  61.38 & 150 MeV \\
150$^+$ MeV&1.74\E{12}& 2.08\E{-5}	&  17.93 &  26.31 &  20.13 &  24.77 & 214 $+$170 MeV \\
		&			  & 1.36\E{-5}	&  54.65 &  61.45 &  60.65 &  61.25 & 150 MeV \\		
150$^-$ MeV&1.74\E{12}& 2.08\E{-5}	&  17.93 &  26.32 &  20.13 &  24.77 & All \\
140 MeV & 1.62\E{12}  & 2.51\E{-5}	&  16.96 &  23.77 &  19.05 &  22.59 & All \\
130 MeV & 1.51\E{12}  & 3.04\E{-5}	&  16.16 &  21.76 &  18.16 &  20.86 & All  \\
100 MeV	& 1.16\E{12}  & 5.66\E{-5}	&  14.39 &  18.00 &  16.21 &  17.55 & All \\
 50 MeV	& 5.80\E{11}  & 2.51\E{-4}  &  11.87 &  14.63 &  13.40 &  14.32 & All\\
 20 MeV	& 2.32\E{11}  & 1.78\E{-3}	&   9.71 &  11.33 &  10.99 &  11.25 & All \\
 10 MeV	& 1.16\E{11}  & 7.32\E{-3}	&   9.50 &  10.76 &  10.75 &  10.76 & All  \\
  5 MeV	& 5.80\E{10}  & 2.93\E{-2}	&   9.49 &  10.74 &  10.73 &  10.74 & All \\
  2 MeV	& 2.32\E{10}  & 0.18		&   9.43 &  10.71 &  10.65 &  10.70 & All  \\
  1 MeV	& 1.16\E{10}  & 0.74		&   9.22 &  10.60 &  10.36 &  10.56 & All  \\
500 keV & 5.80\E{9}	  & 3.11		&   8.53 &  10.16 &   9.43 &  10.03 & All  \\
200 keV & 2.32\E{9}	  & 2.39\E{1}	&   5.97 &   7.66 &   6.20 &   7.55 & All \\
100 keV & 1.16\E{9}	  & 1.22\E{2}	&   4.03 &   4.46 &   3.84 &   4.78 & All \\
 50 keV & 5.80\E{8}	  & 5.24\E{2}	&   3.64 &   3.39 &   3.37 &   3.93 & All \\	
 20 keV & 2.32\E{8}	  & 3.27\E{3}	&   3.64 &   3.36 &	  3.36 &   3.91 & All \\
 10 keV & 1.16\E{8}	  & 1.31\E{4}	&   3.64 &   3.36 &	  3.36 &   3.91 & All  \\
\end{longtable}
\end{footnotesize}

\section{Effective Contribution to One Single Intrinsic Degree of Freedom}
\label{Sec:OneParticleContribution}
\makeatletter
\setcounter{table}{0}
\@addtoreset{table}{section}
\renewcommand{\thetable}{B\arabic{table}}
\makeatletter
\vspace{-6pt}
\begin{table}[H] 	
\footnotesize
\centering
	\caption{The effective contribution to a~single degree of freedom as function of temperature $(\kB T)$ over mass $(mc^2)$.}
	\label{Tab:OneParticleContribution}
	\scalebox{0.94}[0.94]{\begin{tabular}{cccccccccc}
\toprule
	\multirow{2}{*}{\boldmath{$\dfrac{\kB T}{mc^2}$}}& \multicolumn{2}{c}{\textbf{Number Density}}
	& \multicolumn{2}{c}{\textbf{Energy Density}}
	& \multicolumn{2}{c}{\textbf{Pressure}}
	& \multicolumn{2}{c}{\textbf{Entropy Density}} \\ \cmidrule{2-9}
 	& \textbf{Bosons} 	 & \textbf{Fermions}   & \textbf{Bosons}    & \textbf{Fermions}   & \textbf{Bosons}    & \textbf{Fermions}   & \textbf{Bosons}     & \textbf{Fermions} \\
\midrule		
$\infty$& 1		 	  & 0.750	    & 1		      & 0.875	    & 1			  & 0.875	    & 1		      & 0.875	 	\\
10:1	& 0.993		  & 0.749	    & 0.999	      & 0.874		& 0.996		  & 0.873	    & 0.998	  	  & 0.874		\\
2:1		& 0.901		  & 0.716	    & 0.970	      & 0.859		& 0.929		  & 0.831	    & 0.960		  & 0.852		\\
1:1		& 0.740		  & 0.630	    & 0.890	      & 0.808		& 0.784		  & 0.724	    & 0.863		  & 0.787		\\
1:2		& 0.438		  & 0.409	    & 0.658	      & 0.626		& 0.477		  & 0.461	    & 0.613		  & 0.585		\\
1:3		& 0.236		  & 0.227	    & 0.427	      & 0.418		& 0.257		  & 0.254	    & 0.385		  & 0.377		\\
1:4		& 0.116		  & 0.115	    & 0.253	      & 0.251		& 0.129		  & 0.128	    & 0.222		  & 0.222		\\
1:5		& 0.055		  & 0.055	    & 0.139	      & 0.139		& 0.061		  & 0.061	    & 0.120		  & 0.120		\\
1:6		& 0.025		  & 0.025	    & 0.073	      & 0.073		& 0.028		  & 0.028	    & 0.062		  & 0.062		\\
1:7		& 0.011	 	  & 0.011	    & 0.037	      & 0.037		& 0.013	 	  & 0.013	    & 0.031		  & 0.031		\\
1:8		& 4.93\E{-3}  & 4.93\E{-3}	& 0.018	      & 0.018		& 5.48\E{-3}  & 5.48\E{-3}	& 0.015 	  & 0.015		\\
1:9		& 2.12\E{-3}  & 2.12\E{-3}	& 8.37\E{-3}  & 8.37\E{-3}	& 2.35\E{-3}  & 2.35\E{-3}	& 6.87\E{-3}  & 6.87\E{-3}	\\
1:10	& 8.94\E{-4}  & 8.94\E{-4}	& 3.87\E{-3}  & 3.87\E{-3}  & 9.94\E{-4}  & 9.94\E{-4}	& 3.15\E{-3}  & 3.15\E{-3}	\\
1:12	& 1.55\E{-4}  & 1.55\E{-4}	& 7.81\E{-4}  & 7.81\E{-4}	& 1.72\E{-4}  & 1.72\E{-4}	& 6.29\E{-4}  & 6.29\E{-4}	\\
1:14	& 2.58\E{-5}  & 2.58\E{-5}	& 1.49\E{-4}  & 1.49\E{-4}	& 2.87\E{-5}  & 2.87\E{-5}	& 1.19\E{-4}  & 1.19\E{-4}	\\
1:16	& 4.21\E{-6}  & 4.21\E{-6}	& 2.74\E{-5}  & 2.74\E{-5}	& 4.67\E{-6}  & 4.67\E{-6}	& 2.17\E{-5}  & 2.17\E{-5}	\\
1:18	& 6.71\E{-7}  & 6.71\E{-7}	& 4.87\E{-6}  & 4.87\E{-6}	& 7.45\E{-7}  & 7.45\E{-7}	& 3.84\E{-6}  & 3.84\E{-6}	\\
1:20	& 1.05\E{-7}  & 1.05\E{-7}  & 8.42\E{-7}  & 8.42\E{-7}	& 1.17\E{-7}  & 1.17\E{-7}	& 6.61\E{-7}  & 6.61\E{-7}	\\
1:30	& 8.52\E{-12} & 8.52\E{-12} & 9.96\E{-11} & 9.96\E{-11}	& 9.47\E{-12} & 9.47\E{-12} & 7.71\E{-11} & 7.71\E{-11} \\
1:40	& 5.87\E{-16} & 5.87\E{-16} & 9.03\E{-15} & 9.03\E{-15} & 6.52\E{-16} & 6.52\E{-16} & 6.93\E{-15} & 6.93\E{-15} \\
1:50	& 3.69\E{-20} & 3.69\E{-20} & 7.04\E{-19} & 7.04\E{-19} & 4.10\E{-20} & 4.10\E{-20} & 5.38\E{-19} & 5.38\E{-19} \\
1:100	& 1.98\E{-41} & 1.98\E{-41} & 7.43\E{-40} & 7.43\E{-40}	& 2.19\E{-41} & 2.19\E{-41} & 5.62\E{-40} & 5.62\E{-40} \\
\bottomrule
\end{tabular}}
\end{table}

\newpage
\section{List of Mesons and Their Degeneracy} 
\label{Sec:MesonTable}
\vspace{-12pt}
\begin{table}[H]
\small
\centering
	\caption{Pseudoscalar mesons}
\begin{tabular}{cccccc}
\toprule
	\textbf{Symbol} 
	& {\textbf{Flavours}} &{\textbf{Spin States}} & {\textbf{Color States}} & {\textbf{Bose or Fermi}}   & {\textbf{Mass}}  \\
    \midrule
	$\pi^0$       										& 1 & 1 & 1 & 1  	& 134.9766 	\\
	$\pi^{\pm}$											& 2 & 1 & 1 & 1  	& 139.57018\\	
	K$^\pm$												& 2 & 1 & 1 & 1 	& 493.677\\
	K$^0 \bar{\textrm{K}}^0$							& 2 & 1 & 1 & 1		& 497.614\\
	K$^0_\textrm{S}$,  K$^0_\textrm{L}$					& 2 & 1 & 1 & 1		& 497.614\\
	$\eta$												& 1 & 1 & 1 & 1 	& 547.862	\\
	$\eta'$												& 1 & 1 & 1 & 1 	& 957.78	\\
	D$^0$, $\bar{\textrm{D}}^0$							& 2 & 1 & 1 & 1 	& 1864.84\\
	D$^\pm$ 											& 2 & 1 & 1 & 1 	& 1869.61	\\
	D$_\textrm{s}^\pm$									& 2 & 1 & 1 & 1 	& 1968.30	\\
	$\eta_\textrm{c}$									& 1 & 1 & 1 & 1 	& 2983.6	\\
	B$^\pm$												& 2 & 1 & 1 & 1 	& 5279.26	\\
	B$^0$, $\bar{\textrm{B}}^0$							& 2 & 1 & 1 & 1 	& 5279.58	\\
	B$_\textrm{s}^0$, $\bar{\textrm{B}_\textrm{s}}^0$	& 2 & 1 & 1 & 1 	& 5366.77	\\
	B$_\textrm{c}^\pm$									& 2 & 1 & 1 & 1 	& 6275.6 \\
	$\eta_\textrm{b}$									& 1 & 1 & 1 & 1 	& 9398.0	\\	
    \midrule
\multicolumn{5}{l}{\textbf{Pseudoscalar Mesons}} & \multicolumn{1}{c}{$\mathbf{g = 27}$} \\
    \bottomrule
    \end{tabular}
\end{table}
\unskip
  \begin{table}[H]
\small
\centering
	\caption{Vector mesons}
\begin{tabular}{cccccc}
\toprule
   	\textbf{Symbol} 
   	& {\textbf{Flavours}} &{\textbf{Spin States}} & {\textbf{Color States}} & {\textbf{Bose or Fermi}}   & {\textbf{Mass}}  \\
    \midrule
	$\rho^\pm$											& 2 & 3 & 1 & 1 	& 775.11 \\
	$\rho^0$											& 1 & 3 & 1 & 1		& 775.26 \\
	$\omega$											& 1 & 3 & 1 & 1		& 782.65 \\
	K$^{* \pm}$											& 2 & 3 & 1 & 1		& 891.66 \\
	K$^{* 0}$, $\bar{\textrm{K}}^{* 0}$					& 2 & 3 & 1 & 1		& 895.81 \\
	$\phi$												& 1 & 3 & 1 & 1		& 1019.461 \\
	D$^{* 0}$, $\bar{\textrm{D}}^{* 0}$					& 2 & 3 & 1 & 1		& 2006.96 \\
	D$^{* \pm}$											& 2 & 3 & 1 & 1		& 2010.26 \\
	D$_\textrm{s}^{\star \pm}$							& 2 & 3 & 1 & 1		& 2112.1 \\
	J/$\psi$											& 1 & 3 & 1 & 1		& 3096.916 \\
	B$^{* \pm}$											& 2 & 3 & 1 & 1		& 5325.2\\
	B$^{* 0}$, $\bar{\textrm{B}}^{* 0}$					& 2 & 3 & 1 & 1		& 5325.2 $^a$ \\
	B$_\textrm{s}^{* 0}$, $\bar{\textrm{B}}_\textrm{s}^{* 0}$& 2 & 3 & 1 & 1& 5415.4 $^a$ \\
	$\Upsilon$ (1S)										& 1 & 3 & 1 & 1		& 9460.30 \\
    \midrule
\multicolumn{5}{l}{\textbf{Vector Mesons}} & \multicolumn{1}{c}{$\mathbf{g = 69}$} \\
    \bottomrule
\end{tabular}\\
\begin{tabular}{cccccc}
\multicolumn{1}{c}{\footnotesize $^a$ These are listed without 0 superscript by PDG \cite{PDG:2014}.} 
\end{tabular}
\end{table}
\unskip
\begin{table}[H]
\footnotesize
\centering
\caption{Excited unflavoured mesons}
\begin{tabular}{cccccc}
  \toprule
	\textbf{Symbol} 
	& {\textbf{Flavours}} & {\textbf{Spin States}} & {\textbf{Color States}} & {\textbf{Bose or Fermi}}   & \multicolumn{1}{c}{\textbf{Mass}}  \\
  \midrule
	f$_0$(500) 			& 1 & 1 & 1 & 1 & 500	\\
	f$_0$(980) 			& 1 & 1 & 1 & 1 & 980	\\
	a$_0$(980) 			& 3 & 1 & 1 & 1 & 980	\\
	h$_1$(1170)			& 1 & 3 & 1 & 1 & 1170	\\
	b$_1$(1235)			& 3 & 3 & 1 & 1 & 1235	\\
	a$_1$(1260)			& 3 & 3 & 1 & 1 & 1260	\\
	f$_2$(1270)  		& 1 & 5 & 1 & 1 & 1270	\\
	f$_1$(1285)  		& 1 & 3 & 1 & 1 & 1285	\\
	$\eta$(1295)		& 1 & 1 & 1 & 1 & 1295	\\
	$\pi$(1300)			& 3 & 1 & 1 & 1 & 1300	\\
	a$_2$(1320)  		& 3 & 5 & 1 & 1 & 1320	\\
	f$_0$(1370)  		& 1 & 1 & 1 & 1 & 1370	\\
	h$_1$(1380)  		& 1 & 3 & 1 & 1 & 1380	\\
	$\pi_1$(1400)  		& 3 & 3 & 1 & 1 & 1400	\\
	$\eta$(1405)  		& 1 & 1 & 1 & 1 & 1405	\\
	f$_1$(1420)  		& 1 & 3 & 1 & 1 & 1420	\\
	$\omega$(1420)  	& 1 & 3 & 1 & 1 & 1420	\\
	f$_2$(1430)  		& 1 & 5 & 1 & 1 & 1430	\\
	a$_0$(1450)  		& 3 & 1 & 1 & 1 & 1450	\\
	$\rho$(1450) 	 	& 3 & 3 & 1 & 1 & 1450	\\
	$\eta$(1475) 	 	& 1 & 1 & 1 & 1 & 1475	\\
	f$_0$(1500)  		& 1 & 1 & 1 & 1 & 1500	\\
	f$_1$(1510)  		& 1 & 3 & 1 & 1 & 1510	\\
	f'$_1$(1525) 	 	& 1 & 3 & 1 & 1 & 1525	\\
	f$_2$(1565)  		& 1 & 5 & 1 & 1 & 1565\\
	$\rho$(1570) 	 	& 3 & 3 & 1 & 1 & 1570	\\
	h$_1$(1595)  		& 1 & 3 & 1 & 1 & 1595	\\
	$\pi_1$(1600)  		& 3 & 3 & 1 & 1 & 1600	\\
	a$_1$(1640)			& 3 & 3 & 1 & 1 & 1640	\\
	f$_2$(1640)			& 1 & 5 & 1 & 1 & 1640  \\
	$\eta_2$(1645)		& 1 & 5 & 1 & 1 & 1645\\
	$\omega$(1650)		& 1 & 3 & 1 & 1 & 1650  \\
	$\omega_3$(1670)	& 1 & 7 & 1 & 1 & 1670  \\
	$\pi_2$(1670)		& 3 & 5 & 1 & 1 & 1670  \\
	$\phi$(1680)		& 1 & 3 & 1 & 1 & 1680  \\
	$\rho_3$(1690)		& 3 & 7 & 1 & 1 & 1690  \\
	$\rho$(1700)		& 3 & 3 & 1 & 1 & 1700  \\
	a$_2$(1700)			& 3 & 5 & 1 & 1 & 1700  \\
	f$_0$(1710)			& 1 & 1 & 1 & 1 & 1710  \\
	$\eta$(1760)		& 1 & 1 & 1 & 1 & 1760  \\
	$\pi$(1800)			& 3 & 1 & 1 & 1 & 1800  \\
	f$_2$(1810)			& 1 & 5 & 1 & 1 & 1800  \\
	$\phi_3$(1850)		& 1 & 7 & 1 & 1 & 1850  \\
	$\eta_2$(1870)		& 1 & 5 & 1 & 1 & 1870  \\
	$\pi_2$(1880)		& 3 & 5 & 1 & 1 & 1880  \\
	$\rho$(1900)		& 3 & 1 & 1 & 1 & 1900  \\
	f$_2$(1910)			& 1 & 5 & 1 & 1 & 1910  \\
	f$_2$(1950)			& 1 & 5 & 1 & 1 & 1950  \\
	$\rho_3$(1990)		& 3 & 7 & 1 & 1 & 1990  \\
	f$_2$(2010)			& 1 & 5 & 1 & 1 & 2010  \\
	f$_0$(2020)			& 1 & 1 & 1 & 1 & 2020  \\
	a$_4$(2040)			& 3 & 9 & 1 & 1 & 2040  \\
	f$_4$(2050)			& 1 & 9 & 1 & 1 & 2050  \\
	$\pi_2$(2100)		& 3 & 5 & 1 & 1 & 2100  \\
	f$_0$(2100)			& 1 & 1 & 1 & 1 & 2100  \\
	f$_2$(2150)			& 1 & 5 & 1 & 1 & 2150  \\
	$\rho$(2150)		& 3 & 3 & 1 & 1 & 2150  \\
	$\phi$(2170)		& 1 & 3 & 1 & 1 & 2170  \\
	f$_0$(2200)			& 1 & 1 & 1 & 1 & 2200  \\
	f$_J(2220)$ 		& 1 & 9 & 1 & 1 & 2220 $^b$ \\	 
	$\eta$(2225)		& 1 & 1 & 1 & 1 & 2225  \\
	$\rho_3$(2250)		& 3 & 7 & 1 & 1 & 2250  \\
	f$_2$(2300)			& 1 & 5 & 1 & 1 & 2300  \\
	f$_4$(2300)			& 1 & 9 & 1 & 1 & 2300  \\
	f$_0$(2330)			& 1 & 1 & 1 & 1 & 2330  \\
	f$_2$(2340)			& 1 & 5 & 1 & 1 & 2340  \\
	$\rho_5$(2350)		& 3 &11 & 1 & 1 & 2350  \\	
	a$_6$(2450)			& 3 &13 & 1 & 1 & 2450  \\	
	f$_6$(2510)			& 1 &13 & 1 & 1 & 2510  \\	
\midrule
\multicolumn{5}{l}{\textbf{Excited Unflavoured Mesons}} & \multicolumn{1}{r}{$\mathbf{g = 499}$} \\
\bottomrule
\end{tabular}\\
\begin{tabular}{cccccc}
\multicolumn{1}{c}{\footnotesize $^b$ We have used the $(J^{PC}) = (4^{++})$ for f$_J$.}
\end{tabular}
\end{table}
\begin{table}[H]
\small
\centering
\caption{Excited strange mesons}
\begin{tabular}{cccccc}
\toprule
	\textbf{Symbol} 
	&{\textbf{Flavours}} &{\textbf{Spin States}} & {\textbf{Color States}} & {\textbf{Bose or Fermi}}   & \multicolumn{1}{l}{\textbf{Mass}}  \\
  \midrule
	K$_1$(1270)  		& 2 & 3 & 1 & 1 & 1270 \\
	K$_1$(1400)  		& 2 & 3 & 1 & 1 & 1400\\
	K$^*$(1410)  		& 2 & 3 & 1 & 1 & 1410	\\
	K$^*_0$(1430)  		& 2 & 1 & 1 & 1 & 1430	\\
	K$^*_2$(1430)  		& 2 & 5 & 1 & 1 & 1430\\
	K$^*_2$(1430)  		& 2 & 5 & 1 & 1 & 1430	\\
	K(1460)  			& 2 & 1 & 1 & 1 & 1460	\\
	K$_2$(1580)  		& 2 & 5 & 1 & 1 & 1580	\\
	K$_1$(1650)  		& 2 & 3 & 1 & 1 & 1650	\\
	K$^*$(1680)  		& 2 & 3 & 1 & 1 & 1680	\\
	K$_2$(1770)  		& 2 & 5 & 1 & 1 & 1770\\
	K$^*_3$(1780)  		& 2 & 7 & 1 & 1 & 1780	\\
	K$_2$(1820)  		& 2 & 5 & 1 & 1 & 1820	\\
	K(1830)		  		& 2 & 1 & 1 & 1 & 1830	\\
	K$^*_0$(1950)  		& 2 & 1 & 1 & 1 & 1950	\\
	K$^*_2$(1980)  		& 2 & 5 & 1 & 1 & 1980	\\
	K$^*_4$(2045)  		& 2 & 9 & 1 & 1 & 2045	\\
	K$_2$(2250)  		& 2 & 5 & 1 & 1 & 2250	\\
	K$_3$(2320)  		& 2 & 7 & 1 & 1 & 2320	\\
	K$^*_5$(2380)  		& 2 &11 & 1 & 1 & 2380	\\	
	K$_4$(2500)  		& 2 & 9 & 1 & 1 & 2500	\\
    \midrule
\multicolumn{5}{l}{\textbf{Excited Strange Mesons}} & \multicolumn{1}{r}{$\mathbf{g = 194}$} \\
\bottomrule
\end{tabular}
\end{table}
\setlength{\tabcolsep}{0.3em} 
\newcolumntype{C}[1]{%
 >{\vbox to 2.5ex\bgroup\vfill\centering}%
 p{#1}%
 <{\egroup}}

\section{List of Baryons and Their Degeneracy}
\label{Sec:BaryonTable}
\vspace{-12pt}
\begin{table}[H]
\small
\centering
\caption{Spin \oha~ baryons}
\begin{tabular}{cccccc}
\toprule
	\textbf{Symbol} 
	& {\textbf{Flavours}} & {\textbf{Spin States}} & {\textbf{Color States}} & {\textbf{Bose or Fermi}}   & \multicolumn{1}{c}{\textbf{Mass}}  \\
\midrule
	p						& 2 & 2 & 1 & \f 	&  938.272 \\
	n						& 2 & 2 & 1	& \f 	&  939.565    \\
	$\Lambda^0$				& 2 & 2 & 1	& \f 	& 1115.683 	 \\
	$\Sigma^+$				& 2 & 2 & 1	& \f 	& 1189.37	 \\
	$\Sigma^0$				& 2 & 2 & 1	& \f 	& 1192.642	  \\
	$\Sigma^-$				& 2 & 2 & 1	& \f 	& 1197.449	 \\
	$\Xi^0$					& 2 & 2 & 1	& \f 	& 1314.86	\\
	$\Xi^-$					& 2 & 2 & 1	& \f 	& 1321.71	 \\
	$\Lambda^+_\textrm{c}$	& 2 & 2 & 1	& \f 	& 2186.46	  \\
	$\Sigma^+_\textrm{c}$		& 2 & 2 & 1	& \f 	& 2452.9	 \\
	$\Sigma^0_\textrm{c}$		& 2 & 2 & 1	& \f 	& 2453.74	  \\
	$\Sigma^{++}_\textrm{c}$	& 2 & 2 & 1	& \f 	& 2453.98	\\
	$\Xi^{+}_\textrm{c}$		& 2 & 2 & 1	& \f 	& 2467.8	  \\
	$\Xi^{0}_\textrm{c}$		& 2 & 2 & 1	& \f 	& 2470.88	  \\
	$\Xi'^{+}_\textrm{c}$		& 2 & 2 & 1	& \f 	& 2575.6	 \\
	$\Xi'^{0}_\textrm{c}$		& 2 & 2 & 1	& \f 	& 2577.9	 \\
	$\Omega^{0}_\textrm{c}$	& 2 & 2 & 1	& \f 	& 2695.2	 \\
	$\Xi^{+}_\textrm{cc}$		& 2 & 2 & 1	& \f 	& 3518.9	\\
	$\Lambda^0_\textrm{b}$	& 2 & 2 & 1	& \f 	& 5619.4	\\
	$\Xi^0_\textrm{b}$		& 2 & 2 & 1	& \f 	& 5787.8	 \\
	$\Xi^-_\textrm{b}$		& 2 & 2 & 1	& \f 	& 5791.1	  \\
	$\Sigma^+_\textrm{b}$		& 2 & 2 & 1	& \f 	& 5811.3	  \\
	$\Sigma^-_\textrm{b}$		& 2 & 2 & 1	& \f 	& 5815.5	\\
	$\Omega^-_\textrm{b}$		& 2 & 2 & 1	& \f 	& 6071		  \\
    \midrule
\multicolumn{5}{l}{\textbf{Spin \oha~ Baryons}} & \multicolumn{1}{r}{$\mathbf{g = 84}$} \\
\bottomrule
\end{tabular}
\end{table}
\begin{table}[H]
\small
\centering
\caption{Spin \tha~ baryons}
\begin{tabular}{cccccc}
\toprule
   	\textbf{Symbol} 
   	& {\textbf{Flavours}} & {\textbf{Spin States}} & {\textbf{Color States}} & {\textbf{Bose or Fermi}}   & \multicolumn{1}{c}{\textbf{Mass}}  \\
    \midrule
	$\Delta^{++}$			& 2 & 4 & 1 & \f 	& 1232    \\
	$\Delta^{+}$			& 2 & 4 & 1 & \f 	& 1232   \\
	$\Delta^{0}$			& 2 & 4 & 1 & \f 	& 1232   \\
	$\Delta^{-}$			& 2 & 4 & 1 & \f 	& 1232    \\
	$\Sigma^{*+}$			& 2 & 4 & 1 & \f 	& 1382.8  \\
	$\Sigma^{*0}$			& 2 & 4 & 1 & \f 	& 1383.7  \\
	$\Sigma^{*-}$			& 2 & 4 & 1 & \f 	& 1387.2  \\
	$\Xi^{*0}$				& 2 & 4 & 1 & \f 	& 1531.80 \\
	$\Xi^{*-}$				& 2 & 4 & 1 & \f 	& 1535.0  \\
	$\Omega^{-}$			& 2 & 4 & 1 & \f 	& 1672.45 \\
	$\Sigma^{*+}_\textrm{c}$	& 2 & 4 & 1 & \f 	& 2517.5  \\
	$\Sigma^{*++}_\textrm{c}$	& 2 & 4 & 1 & \f 	& 2517.9  \\
	$\Sigma^{*0}_\textrm{c}$	& 2 & 4 & 1 & \f 	& 2518.8  \\
	$\Xi^{*+}_\textrm{c}$		& 2 & 4 & 1 & \f 	& 2645.9 \\
	$\Xi^{*0}_\textrm{c}$		& 2 & 4 & 1 & \f 	& 2645.9  \\
	$\Omega^{*0}_\textrm{c}$	& 2 & 4 & 1 & \f 	& 2765.9 \\
	$\Sigma^{*+}_\textrm{b}$	& 2 & 4 & 1 & \f 	& 5832.1  \\
	$\Sigma^{*-}_\textrm{b}$	& 2 & 4 & 1 & \f 	& 5835.1 \\
	$\Xi^{*-}_\textrm{b}$		& 2 & 4 & 1 & \f 	& 5945.5  \\
 \midrule
\multicolumn{5}{l}{\textbf{Spin \tha~Baryons}} & \multicolumn{1}{r}{$\mathbf{g = 133}$}  \\
\bottomrule
\end{tabular}
\end{table}
\unskip
\begin{table}[H]
\small
\centering
\caption{Excited N baryons}
\begin{tabular}{cccccc}
\toprule
\textbf{Symbol} 
& {\textbf{Flavours}} & {\textbf{Spin States}} & {\textbf{Color States}} & {\textbf{Bose or Fermi}}   & \multicolumn{1}{c}{\textbf{Mass}}  \\
 \midrule
	N(1440)					& 2 & 2 & 1 & \f  & 1440  \\
	N(1520)					& 2 & 4 & 1 & \f  & 1520  \\	
	N(1535)					& 2 & 2 & 1 & \f  & 1535 \\	
	N(1650)					& 2 & 2 & 1 & \f  & 1650  \\	
	N(1675)					& 2 & 6 & 1 & \f  & 1675  \\	
	N(1680)					& 2 & 6 & 1 & \f  & 1680 \\
	N(1700)   				& 2 & 4 & 1 & \f  & 1700  \\
	N(1710)   				& 2 & 2 & 1 & \f  & 1710  \\
	N(1720)   				& 2 & 4 & 1 & \f  & 1720  \\	
	N(1860)   				& 2 & 6 & 1 & \f  & 1860  \\	
	N(1875)   				& 2 & 4 & 1 & \f  & 1875  \\	
	N(1880)   				& 2 & 2 & 1 & \f  & 1880  \\
	N(1895)   				& 2 & 2 & 1 & \f  & 1895  \\	
	N(1900)   				& 2 & 4 & 1 & \f  & 1900 \\	
	N(1990)   				& 2 & 8 & 1 & \f  & 1990  \\	
	N(2000)   				& 2 & 6 & 1 & \f  & 2000 \\	
	N(2040)   				& 2 & 4 & 1 & \f  & 2040  \\	
	N(2060)   				& 2 & 6 & 1 & \f  & 2060  \\	
	N(2100)   				& 2 & 2 & 1 & \f  & 2100  \\	
	N(2120)   				& 2 & 4 & 1 & \f  & 2120  \\	
	N(2190)   				& 2 & 8 & 1 & \f  & 2190  \\	
	N(2220)   				& 2 &10 & 1 & \f  & 2220  \\
	N(2250)   				& 2 &10 & 1 & \f  & 2250  \\	
	N(2300)   				& 2 & 2 & 1 & \f  & 2300  \\	
	N(2570)   				& 2 & 6 & 1 & \f  & 2570  \\	
	N(2600)   				& 2 &12 & 1 & \f  & 2600  \\	
	N(2700)   				& 2 &14 & 1 & \f  & 2700  \\
\midrule
\multicolumn{5}{l}{\textbf{Excited N Baryons}} & \multicolumn{1}{r}{$\mathbf{g = 248.5}$}   \\
\bottomrule
\end{tabular}
\end{table}
\begin{table}[H]
\small
\centering
\caption{Excited $\Delta$ Baryons}
\begin{tabular}{cccccc}
\toprule
\textbf{Symbol} 
& {\textbf{Flavours}} & {\textbf{Spin States}} & {\textbf{Color States}} & {\textbf{Bose or Fermi}}   & \multicolumn{1}{c}{\textbf{Mass}}  \\
    \midrule
	$\Delta$(1600)			& 2 & 4 & 1 & \f  & 1600  \\	
	$\Delta$(1620)			& 2 & 2 & 1 & \f  & 1620  \\	
	$\Delta$(1700)			& 2 & 4 & 1 & \f  & 1700  \\
	$\Delta$(1750)			& 2 & 2 & 1 & \f  & 1750  \\
	$\Delta$(1900)			& 2 & 2 & 1 & \f  & 1900  \\
	$\Delta$(1905)			& 2 & 6 & 1 & \f  & 1905  \\
	$\Delta$(1910)			& 2 & 2 & 1 & \f  & 1910  \\
	$\Delta$(1920)			& 2 & 4 & 1 & \f  & 1920  \\
	$\Delta$(1930)			& 2 & 6 & 1 & \f  & 1930  \\
	$\Delta$(1940)			& 2 & 4 & 1 & \f  & 1940  \\
	$\Delta$(1950)			& 2 & 8 & 1 & \f  & 1950  \\
	$\Delta$(2000)			& 2 & 6 & 1 & \f  & 2000  \\
	$\Delta$(2150)			& 2 & 2 & 1 & \f  & 2150  \\
	$\Delta$(2200)			& 2 & 8 & 1 & \f  & 2200  \\
	$\Delta$(2300)			& 2 &10 & 1 & \f  & 2300  \\
	$\Delta$(2350)			& 2 & 6 & 1 & \f  & 2350  \\
	$\Delta$(2390)			& 2 & 8 & 1 & \f  & 2390  \\
	$\Delta$(2400)			& 2 &10 & 1 & \f  & 2400  \\
	$\Delta$(2420)			& 2 &12 & 1 & \f  & 2420  \\
	$\Delta$(2750)			& 2 &14 & 1 & \f  & 2750  \\
	$\Delta$(2950)			& 2 &16 & 1 & \f  & 2950  \\
    \midrule
\multicolumn{5}{l}{\textbf{Excited $\Delta$ Baryons}} & \multicolumn{1}{r}{$\mathbf{g = 238}$}  \\
\bottomrule
\end{tabular}
\end{table}
\unskip
\begin{table}[H]
\small
\centering
\caption{Excited $\Lambda$ baryons}
\begin{tabular}{cccccc}
\toprule
\textbf{Symbol} 
& {\textbf{Flavours}} & {\textbf{Spin States}} & {\textbf{Color States}} & {\textbf{Bose or Fermi}}   & \multicolumn{1}{c}{\textbf{Mass}}  \\
    \midrule
	$\Lambda$(1405) 		& 2 & 2 & 1 & \f  & 1405 \\
	$\Lambda$(1520) 		& 2 & 4 & 1 & \f  & 1520  \\	
	$\Lambda$(1600)  		& 2 & 2 & 1 & \f  & 1600  \\	
	$\Lambda$(1670)  		& 2 & 2 & 1 & \f  & 1670  \\	
	$\Lambda$(1690)  		& 2 & 4 & 1 & \f  & 1690  \\
	$\Lambda$(1710)  		& 2 & 2 & 1 & \f  & 1710  \\
	$\Lambda$(1800)  		& 2 & 2 & 1 & \f  & 1800  \\
	$\Lambda$(1810)  		& 2 & 2 & 1 & \f  & 1810  \\
	$\Lambda$(1820)  		& 2 & 6 & 1 & \f  & 1820  \\
	$\Lambda$(1830)  		& 2 & 6 & 1 & \f  & 1830  \\
	$\Lambda$(1890)  		& 2 & 4 & 1 & \f  & 1890  \\
	$\Lambda$(2020)  		& 2 & 8 & 1 & \f  & 2020  \\
	$\Lambda$(2050)  		& 2 & 4 & 1 & \f  & 2050  \\
	$\Lambda$(2100)  		& 2 & 8 & 1 & \f  & 2100  \\
	$\Lambda$(2110)  		& 2 & 6 & 1 & \f  & 2110  \\
	$\Lambda$(2325)  		& 2 & 4 & 1 & \f  & 2325  \\
	$\Lambda$(2350)  		& 2 &10 & 1 & \f  & 2350\\
    \midrule
\multicolumn{5}{l}{\textbf{Excited $\Lambda$ Baryons}} & \multicolumn{1}{r}{$\mathbf{g = 133}$}\\
\bottomrule
\end{tabular}
\end{table}
\unskip
\begin{table}[H]
\small
\centering
\caption{Excited $\Sigma$ baryons}
\begin{tabular}{cccccc}
\toprule
	\textbf{Symbol} 
	&{\textbf{Flavours}} & {\textbf{Spin States}} & {\textbf{Color States}} & {\textbf{Bose or Fermi}}   & \multicolumn{1}{c}{\textbf{Mass}}  \\
    \midrule
	$\Sigma$(1580)  		& 2 & 4 & 1 & \f  & 1580  \\
	$\Sigma$(1620)  		& 2 & 2 & 1 & \f  & 1620  \\	
	$\Sigma$(1660)  		& 2 & 2 & 1 & \f  & 1660  \\	
	$\Sigma$(1670)  		& 2 & 4 & 1 & \f  & 1670  \\
	$\Sigma$(1730)  		& 2 & 4 & 1 & \f  & 1730  \\
	$\Sigma$(1750)  		& 2 & 2 & 1 & \f  & 1750  \\
	$\Sigma$(1770)  		& 2 & 2 & 1 & \f  & 1770  \\
	$\Sigma$(1775)  		& 2 & 6 & 1 & \f  & 1775  \\
	$\Sigma$(1840)  		& 2 & 4 & 1 & \f  & 1840  \\
	$\Sigma$(1880)  		& 2 & 2 & 1 & \f  & 1880  \\
	$\Sigma$(1900)  		& 2 & 2 & 1 & \f  & 1900  \\
	$\Sigma$(1915)  		& 2 & 6 & 1 & \f  & 1915  \\
	$\Sigma$(1940$^+$)  	& 2 & 4 & 1 & \f  & 1940  \\
	$\Sigma$(1940$^-$)  	& 2 & 4 & 1 & \f  & 1940  \\
	$\Sigma$(2000)  		& 2 & 2 & 1 & \f  & 2000  \\
	$\Sigma$(2030)  		& 2 & 8 & 1 & \f  & 2030 \\
	$\Sigma$(2070)  		& 2 & 6 & 1 & \f  & 2070  \\
	$\Sigma$(2080)  		& 2 & 4 & 1 & \f  & 2080  \\
	$\Sigma$(2100)  		& 2 & 8 & 1 & \f  & 2100\\
    \midrule
\multicolumn{5}{l}{\textbf{Excited $\Sigma$ Baryons}} & \multicolumn{1}{r}{$\mathbf{g = 133}$}\\
\bottomrule
\end{tabular}
\end{table}
\unskip
\begin{table}[H]
\small
\centering
\caption{Excited $\Xi$ baryons}
\begin{tabular}{cccccc}
\toprule
\textbf{Symbol} 
& {\textbf{Flavours}} & {\textbf{Spin States}} & {\textbf{Color States}} & {\textbf{Bose or Fermi}}   & \multicolumn{1}{c}{\textbf{Mass}}  \\
    \midrule
	$\Xi$(1820) 			& 2 & 4 & 1 & \f  & 1820  \\
	$\Xi$(2030) 			& 2 & 6 & 1 & \f  & 2030 \\
    \midrule
\multicolumn{5}{l}{\textbf{Excited $\Xi$ Baryons}} & \multicolumn{1}{r}{$\mathbf{g = 17.5}$}\\
\bottomrule
\end{tabular}
\end{table}

\vspace{6pt}
\acknowledgments{The author would like to thank Jens O. Andersen, Iver Brevik, K{\aa}re Olaussen, and Alireza~Qaiumzadeh for fruitful discussions.}
\conflictofinterests{The authors declare no conflict of interest.} 

\bibliographystyle{mdpi}
\renewcommand\bibname{References}

\end{document}